\pdfoutput=1
\documentclass[PAPER, atlasdraft=false, UKenglish, cernpreprint, texmf, orcidlogo]{atlasdoc}
\usepackage{atlaspackage}
\usepackage{comment}
\usepackage{pdflscape}
\usepackage{atlasbiblatex}

\usepackage[process,snippets]{atlasphysics}
\graphicspath{{logos/}{figures/}}

\usepackage{ANA-EXOT-2022-12-PAPER-defs}

\usepackage{cleveref}
\usepackage{multirow}
\usepackage{mathtools}
\usepackage[compat=1.1.0]{tikz-feynman}
\tikzfeynmanset{warn luatex=false}

\AtlasTitle{Search for heavy neutral leptons in decays of W bosons using leptonic and semi-leptonic displaced vertices in $\sqrt{s} = 13$ TeV $pp$ collisions with the ATLAS detector}

\AtlasAbstract{%
A search is performed for long-lived heavy neutral leptons (HNLs), produced through the decay of a $W$ boson along with a muon or electron.
Two channels are explored: a leptonic channel, in which the HNL decays into two leptons and a neutrino, and a semi-leptonic channel, in which the HNL decays into a lepton and a charged pion.
The search is performed with 140~fb$^{-1}$ of $\sqrt{s} = 13$ TeV proton--proton collision data collected by ATLAS during Run 2 of the Large Hadron Collider.
No excess of events is observed; Dirac-like and Majorana-like HNLs with masses below 14.5 GeV and mixing coefficients as small as 10$^{-7}$ are excluded at the 95\% confidence level.
The results are interpreted under different assumptions on the flavour of the leptons from the HNL decays.
}

\AtlasRefCode{EXOT-2022-12}

\PreprintIdNumber{CERN-EP-2025-052}

\AtlasJournalRef{\JHEP 07 (2025) 196}
\AtlasDOI{DOI:10.1007/JHEP07(2025)196}

\hypersetup{pdftitle={ATLAS document},pdfauthor={The ATLAS Collaboration}}

\begin{document}

\maketitle

\tableofcontents

\section{Introduction}
\label{sec:intro}

%
In the Standard Model (SM) neutrinos are massless and flavour-conserving.
The observation of neutrino mass through flavour oscillation~\cite{deSalas:2017kay} constitutes a distinct deviation from the SM\@.
The type-I seesaw model accounts for these phenomena by introducing right-handed neutrino states that carry no SM gauge charges, thus allowing them to have Majorana masses~\cite{Minkowski:1977sc,Yanagida:1980xy,Glashow:1979nm,Gell-Mann:1979vob,Mohapatra:1979ia,Schechter:1980gr,Schechter:1981cv}, in which the Majorana mass term implies the particle is its own anti-particle.
The model explains the light neutrino masses and predicts the existence of heavy mass eigenstates known as ``heavy neutral leptons'' (HNLs, also denoted \myhnl).
The HNLs can also account for baryon asymmetry via leptogenesis~\cite{Davidson:2008bu,Pilaftsis:2009pk,Shaposhnikov:2009zzb}, and models with three HNLs can provide a dark matter candidate~\cite{Asaka:2005an,Asaka:2005pn,Boyarsky:2018tvu,Ghiglieri:2020ulj}.
The HNL states experience mixing with SM left-handed neutrinos allowing them to undergo weak interactions.
HNLs have been searched for in a wide range of mixing, mass, and lifetime scenarios~\cite{EXOT-2017-26,Belle:2013ytx,NuTeV:1999kej,DELPHI:1996qcc,CHARMII:1994jjr,NA3:1986ahv,CHARM:1985nku,WA66:1985mfx,CMS-EXO-20-009,CMS-EXO-21-013,EXOT-2019-29,CMS-EXO-21-011,CMS-EXO-22-019,CMS-EXO-22-017,EXOT-2017-26,LHCb-HNL-2021,CMS-EXO-21-003,EXOT-2020-06,EXOT-2023-16,CMS-EXO-22-011}.

The search presented in this paper considers both a simplified model in which the HNL mixes with only one lepton flavour (1SFH) as well as a more complete model with two quasi-degenerate HNLs (2QDH) in which the mixing coefficients to all three lepton generations are non-zero.
For the 2QDH scenarios, a different interpretation is considered for the two neutrino-mass hierarchy scenarios.
In the inverted mass hierarchy case (IH), the relative mixing angles are taken to be $x_{\alpha} \equiv |U_{\alpha}|^2/|U_{\mathrm{tot}}|^2$ = 1/3; for the normal mass hierarchy case (NH), the values $x_e$ = 0.06, $x_\mu$ = 0.48 and $x_\tau$ = 0.46 are used~\cite{JL_benchmark,Agrawal:2021dbo}. Here $|U|^2$ is the coupling of the HNL to the SM, and $|U_{\mathrm{tot}}|^2$ is inversely proportional to the proper-lifetime, $\tau_{\myhnl}$, and the mass $m_{\myhnl}$~\cite{Gronau:1984ct} as follows:
\begin{equation}
|U_{\mathrm{tot}}|^2 \sim \frac{1}{\tau_{\myhnl} m_{\myhnl}^5}.
\label{eq:coupling_proportional}
\end{equation}
The search presented in this paper focuses on long-lived signatures and accesses lower coupling values by probing HNLs with $m_{\myhnl} <$ 20~\GeV.
Prior searches have set 95\% confidence level (CL) exclusion limits on the coupling values of long-lived HNLs to SM leptons in the range of $1 \times 10^{-7}$ -- $7 \times 10^{-5}$ for HNLs with masses of 1 -- 14~\GeV~\cite{CMS-EXO-21-011,CMS-EXO-22-017}.
Limits have been set at higher masses by searches for promptly decaying HNLs but at larger coupling values~\cite{EXOT-2017-26,LHCb-HNL-2021,CMS-EXO-21-003,EXOT-2020-06,EXOT-2023-16,CMS-EXO-22-011}.

\begin{figure}[ht!]
\subfloat[]{  \includegraphics[width=0.42\textwidth]{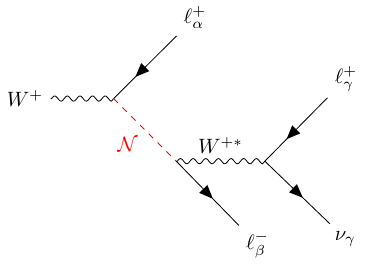} \label{subfig:feyn_lep_W}
} \hfill  \subfloat[]{  \includegraphics[width=0.42\textwidth]{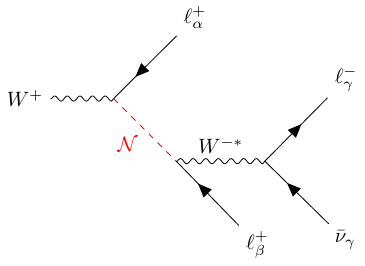} \label{subfig:feyn_lep_W_LNV}
} \hfill
\subfloat[]{  \includegraphics[width=0.42\textwidth]{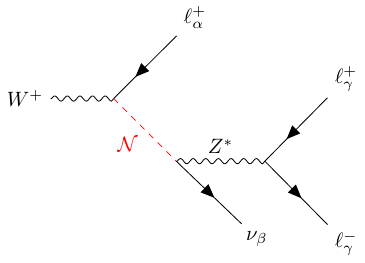} \label{subfig:feyn_lep_Z}
} \hfill  \subfloat[]{  \includegraphics[width=0.42\textwidth]{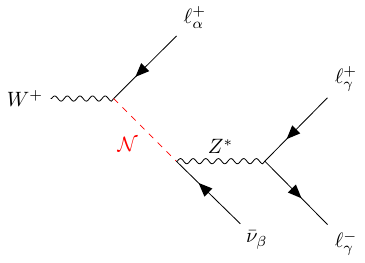} \label{subfig:feyn_lep_Z_LNV}
} \hfill
\subfloat[]{  \includegraphics[width=0.42\textwidth]{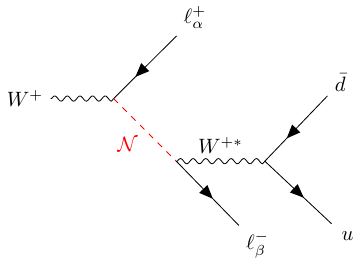}
\label{subfig:feyn_semilep_LNC}
} \hfill   \subfloat[]{  \includegraphics[width=0.42\textwidth]{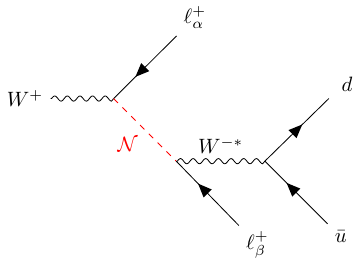}
\label{subfig:feyn_semilep_LNV}
}
\caption{Feynman diagrams for the HNL production and decay processes are shown for both (a) -- (d) leptonic and (e), (f) semi-leptonic decays, in which $\alpha$, $\beta$, $\gamma$ may be $e$ or $\mu$.
Both (a, c, e) LNC, and (b, d, f) LNV processes are shown.
Equivalent diagrams with an initial $W^-$ boson are also considered.
}
\label{fig:feyn}
\end{figure}

The signature studied here is a prompt lepton ($e$, $\mu$), produced near the proton--proton interaction point (IP) from the decay of the $W$ boson, and a displaced two-track vertex (DV), produced far from the IP\@.
In the case that the HNL decays leptonically ($\mathcal{N} \rightarrow \ell\ell\nu$), the DV will consist of two leptons (\ee, $e^{\pm}\mu^{\mp}$, \mumu) as shown in Figure~\ref{subfig:feyn_lep_W}-\ref{subfig:feyn_lep_Z_LNV}.
For the semi-leptonic decays, only decays to a single charged pion are considered, ($\mathcal{N} \rightarrow \ell\pi$), and the DV consists of a lepton and a charged pion ($e^{\pm}\pi^{\mp}$, $\mu^{\pm}\pi^{\mp}$), as shown in Figure~\ref{subfig:feyn_semilep_LNC}-\ref{subfig:feyn_semilep_LNV}.
Other semi-leptonic signal processes are not considered in this search.
If the HNL decays to more than two charged particles, the resulting decay vertex will fail the analysis pre-selection as discussed in Section~\ref{sec:selection}.
If the HNL decays to additional neutral particles, the un-accounted for missing energy causes the event to fail the signal region requirements.
Therefore, other semi-leptonic decays could potentially have small contributions in the studied phase space but the impact to the exclusion contours would be minor.
Likewise, decays to $\tau$-leptons are not included in this analysis. The light decays of $\tau$-leptons to leptons or hadrons are not expected to satisfy the signal selection requirements outlined in Section~\ref{sec:selection}.
Scenarios in which the HNL is either a Dirac fermion (lepton number conserving (LNC)) or a Majorana particle (both LNC and lepton number violating (LNV)) are considered. Due to the mixing of the \myhnl with all three lepton flavours, LNC does not imply lepton flavour conservation.

Events in this search are selected using a single-lepton trigger, making use of the prompt lepton from the $W$-boson decay.
The displaced decay products from the long-lived HNL are reconstructed using large radius tracking (LRT)~\cite{IDTR-2021-03} and a customised displaced vertex reconstruction algorithm.
The two sources of background affecting all channels of the search come from metastable heavy-flavour hadron decays, and from hadrons that are mis-reconstructed as leptons (referred to as the \textit{fake-lepton} background).
Background estimates are performed using a combination of Monte Carlo--driven and data-driven methods.

A significantly improved version of the LRT reconstruction with a large reduction in mis-reconstructed tracks implemented in 2022~\cite{IDTR-2021-03} both reduced the number of background vertices formed by the crossing of unrelated tracks, as well as the CPU usage. This allowed LRT to be run by default in all ATLAS reconstruction, including all Monte Carlo (MC) simulation background samples, in turn making it possible to use MC simulation to directly estimate the background contribution.
With the reduction of mis-reconstructed tracks, it was also possible to include a semi-leptonic channel, in which the two-track vertices have only one track identified as a lepton, without also needing strict selections on the non-leptonic tracks.
The improvement of the LRT reconstruction, in addition to refined analysis techniques, allow for an improved sensitivity compared with the previous ATLAS displaced HNL search~\cite{EXOT-2019-29} using the same dataset.

%
\section{ATLAS detector}
\label{sec:detector}

\newcommand{\AtlasCoordFootnote}{%
ATLAS uses a right-handed coordinate system with its origin at the nominal interaction point (IP)
in the centre of the detector and the \(z\)-axis along the beam pipe.
The \(x\)-axis points from the IP to the centre of the LHC ring,
and the \(y\)-axis points upwards.
Polar coordinates \((r,\phi)\) are used in the transverse plane,
\(\phi\) being the azimuthal angle around the \(z\)-axis.
The pseudorapidity is defined in terms of the polar angle \(\theta\) as \(\eta = -\ln \tan(\theta/2)\) and is equal to the rapidity
$ y = \frac{1}{2} \ln \left( \frac{E + p_z}{E - p_z} \right) $ in the relativistic limit.
Angular distance is measured in units of \(\Delta R \equiv \sqrt{(\Delta y)^{2} + (\Delta\phi)^{2}}\).}

The ATLAS detector~\cite{PERF-2007-01} at the Large Hadron Collider (LHC)~\cite{Evans:2008zzb} covers nearly the entire solid angle around the collision point.\footnote{\AtlasCoordFootnote}
It consists of an inner tracking detector surrounded by a thin superconducting solenoid, electromagnetic and hadronic calorimeters,
and a muon spectrometer incorporating three large superconducting air-core toroidal magnets.

The inner-detector system (ID) is immersed in a \qty{2}{\tesla} axial magnetic field
and provides charged-particle tracking in the range \(|\eta| < 2.5\).
The high-granularity silicon pixel detector covers the vertex region and typically provides four measurements per track, the first hit generally being at $r$ = 33~mm in the insertable B-layer installed before Run~2~\cite{ATLAS-TDR-19,PIX-2018-001}, in addition to the other concentric barrel layers at ${r = 50.5, \, 88.5, \, \mathrm{and} \, 122.5 \, \mathrm{mm}}$, within $|z| <$ 400.5 mm.
The pixel detector has three disks in each of the endcaps at
${|z| = 495, \, 580, \, \mathrm{and} \, 650 \, \mathrm{mm}}$.
It is followed by the SemiConductor Tracker (SCT), which usually provides four pairs of measurements per track, with barrel
layers at $r$ = 299, 371, 443, and 514~mm, within $|z| <$ 746 mm, and nine wheels in each of the endcaps in the range 854 $< |z| <$ 2720 mm.
These silicon detectors are complemented by the transition radiation tracker (TRT),
which enables radially extended track reconstruction up to \(|\eta| = 2.0\).
The TRT also provides electron identification information
based on the fraction of hits (typically 30 in total) above a higher energy-deposit threshold corresponding to transition radiation.

The calorimeter system covers the pseudorapidity range \(|\eta| < 4.9\).
Within the region \(|\eta|< 3.2\), electromagnetic calorimetry is provided by barrel and
endcap high-granularity lead/liquid-argon (LAr) calorimeters,
with an additional thin LAr presampler covering \(|\eta| < 1.8\)
to correct for energy loss in material upstream of the calorimeters.
Hadronic calorimetry is provided by the steel/scintillator-tile calorimeter,
segmented into three barrel structures within \(|\eta| < 1.7\), and two copper/LAr hadronic endcap calorimeters.
The solid angle coverage is completed with forward copper/LAr and tungsten/LAr calorimeter modules
optimised for electromagnetic and hadronic energy measurements respectively.

The muon spectrometer (MS) comprises separate trigger and
high-precision tracking chambers measuring the deflection of muons in a magnetic field generated by the superconducting air-core toroidal magnets.
The field integral of the toroids ranges between \num{2.0} and \qty{6.0}{\tesla\metre}
across most of the detector.
Three layers of precision chambers, each consisting of layers of monitored drift tubes, cover the region \(|\eta| < 2.7\),
complemented by cathode-strip chambers in the forward region, where the background is highest.
The muon trigger system covers the range \(|\eta| < 2.4\) with resistive-plate chambers in the barrel, and thin-gap chambers in the endcap regions.

The luminosity is measured mainly by the LUCID--2~\cite{LUCID2} detector that records Cherenkov light produced in the quartz windows of photomultipliers located close to the beampipe.

Events are selected by the first-level trigger system implemented in custom hardware,
followed by selections made by algorithms implemented in software in the high-level trigger~\cite{TRIG-2016-01}.
The first-level trigger accepts events from the \qty{40}{\MHz} bunch crossings at a rate below \qty{100}{\kHz},
which the high-level trigger further reduces in order to record complete events to disk at about \qty{1}{\kHz}.

A software suite~\cite{SOFT-2022-02} is used in data simulation, in the reconstruction
and analysis of real and simulated data, in detector operations, and in the trigger and data acquisition
systems of the experiment.


%
\section{Data and simulation samples}
\label{sec:dataMC}

This search uses 140 fb$^{-1}$ of proton--proton ($pp$) collision data at $\sqrt{s} = 13$ \TeV, collected by the ATLAS detector at the LHC during the period 2015–2018.
Only data collected when all detector systems were operating acceptably~\cite{DAPR-2018-01} and when the LHC had stable beam conditions are considered.

Monte Carlo samples are used to study the acceptance of the event selection and the efficiency of event reconstruction,
as well as to estimate the background from metastable hadron decays.
The SM background samples are processed with the full ATLAS detector simulation~\cite{SOFT-2010-01} based on $\GEANT$~\cite{Agostinelli:2002hh}.
Simulated signal samples are processed with a fast simulation~\cite{SIMU-2018-04} using a parametrisation of the calorimeter response and $\GEANT$ for the response of the other parts of the detector.
The MC simulated events are reconstructed and analysed with the same analysis chain as data, including the trigger and event selection criteria.
The effects of multiple proton–proton interactions in the same or neighbouring bunch-crossings (pileup) are taken into account by overlaying the simulated hard-scattering process with simulated minimum-bias events.
These events have a mean number of interactions distributed according to their frequency as observed in data, and are generated with \PYTHIA[8.186]~\cite{Sjostrand:2014zea} using the \NNPDF[2.3lo] parton distribution function~(PDF) set~\cite{Ball:2012cx} and the A3 set of tuned parameters (tune)~\cite{ATL-PHYS-PUB-2016-017}. The simulation is adjusted with correction factors to account for differences between data and simulation in the pileup modelling and in the reconstruction and identification efficiencies, and energy and momentum scales of physics objects.

\subsection{Signal simulation}
\label{sec:signal:mc}
Samples of HNL production from $W$-boson decays with leptonic decays of the HNL are generated at next-to-leading-order (NLO) in quantum chromodynamics (QCD) accuracy in \MGNLO[3.3.1]~\cite{Alwall:2014hca} and the \textsc{SM HeavyN NLO} model libraries~\cite{PhysRevD.103.015022}. In these samples, the \NNPDF[3.0nlo] set of PDFs~\cite{Ball:2014uwa} is used.
Samples with semi-leptonic decays of the HNL are generated at leading-order (LO) in QCD using \MADGRAPH[3.5.1] and the \textsc{SM HeavyN Meson NLO} model libraries, using the \NNPDF[3.0nlo\_nf4] set of PDFs. The leptonic (semi-leptonic) signal events are matched with \PYTHIA[8.307]~\cite{Sjostrand:2014zea} (\PYTHIA[8.309]) to model the parton shower, hadronisation, and underlying event, with parameters set according to the A14 tune~\cite{ATL-PHYS-PUB-2014-021} and using the \NNPDF[2.3lo] set of PDFs.
The samples are interfaced to \EVTGEN[2.1.1]~\cite{Lange:2001uf} to model the decays of bottom and charm hadrons.

The HNL is set to decay leptonically or semi-leptonically; the decay kinematics are handled by \MADSPIN~\cite{Frixione:2007zp,Artoisenet:2012st} in the former case, and by \MADGRAPH in the latter.
The process is modelled as a weak decay with a V-A matrix element, and can be mediated by either a $W^*$ boson or, for leptonic decays only, a $Z^*$ boson.
If the charged leptons produced in the HNL decay have the same flavour, then both diagrams (Figures~\ref{subfig:feyn_lep_W} and~\ref{subfig:feyn_lep_W_LNV} or~\ref{subfig:feyn_lep_Z} and~\ref{subfig:feyn_lep_Z_LNV}) contribute to the decay.

For the leptonic decay signal model, samples are simulated for a range of possible HNL masses between 1~\GeV and 20~\GeV for six different final states involving electrons and muons: $\ell-\mu\mu$, $\ell-\mu e$ and $\ell-ee$, with $\ell = e,\mu$. Final states involving $\tau$ leptons are checked to have negligible acceptance.
For the semi-leptonic decay model, the branching fraction of the HNL into a lepton and a pion drastically drops for masses above 3~\GeV~\cite{Bondarenko:2018ptm}; signal samples generated only for $m_\mathcal{N}$ of 1~\GeV, 2~\GeV, and 3~\GeV are investigated. Four separate samples with different final states are generated: $\ell-e\pi$, $\ell-\mu\pi$, with $\ell = e,\mu$. For both the leptonic and semi-leptonic scenarios, mean proper lifetimes of the HNL, $c\tau_{\myhnl}$, are generated between \SI{0.1}{\mm} and \SI{1}{\m}.
For all cases, only one single HNL is generated, in the 1:1:1 coupling scenario, and reweighting is performed to account for the different proposed scenarios.

The total coupling strength of the HNL, $|U_\mathrm{tot}|^2 = \sum_\alpha |U_\alpha|^2 $ in which $\alpha=e,\mu,\tau$ is any given neutrino flavour, can be calculated from the mixing angles that connect the flavour eigenstates of the SM neutrinos with the new heavy HNL mass eigenstates~\cite{JL_benchmark}.
For a Majorana HNL particle  $|U_\mathrm{tot}|^2$ is computed for each of the generated mass and lifetime samples using a direct Fermi theory computation described in Ref.~\cite{Bondarenko:2018ptm}.
The cross-section for the production of a single HNL followed by its decay into the final state particles is calculated for the Majorana model from the leptonic $W$-boson branching ratio $BR(W \rightarrow \ell_\alpha \nu_\alpha)$ according to~\cite{Gonzalez-Garcia:1990sbd}:
\begin{equation}
\begin{aligned}
\sigma_M &= \sigma(pp\rightarrow W) \cdot BR(W \rightarrow \ell_\alpha \myhnl) = \\
&= \sigma(pp\rightarrow W) \cdot BR(W \rightarrow \ell_\alpha \nu_\alpha) \cdot |U_{\alpha}|^2\cdot\left(1- \frac{m_\myhnl^2}{m_W^2}\right)^2 \left( 1 + \frac{m_\myhnl^2}{2m_W^2} \right),
\end{aligned} \label{eq:cross_section}
\end{equation}
where the product of the total $W$-boson production cross-section in 13~\TeV $pp$ collisions times the branching ratio for a $W$-boson decay into a single charged lepton, for $\ell$ = $\mu,e$, is taken from the ATLAS measurement~\cite{STDM-2015-03} to be 20.6 $\pm$ 0.6 nb.
The mass of the $W$ boson is represented by $m_{W}$.

The relationship between the production cross-section for a single Majorana HNL, $\sigma_M$, and the cross-sections for the 1SFH and 2QDH models are shown in Table~\ref{tab:xsecs:models}.
The Dirac and Majorana limits refer to scenarios in which only LNC, or both LNC and LNV processes can occur, respectively. In the context of the two-HNL model, these limits correspond to specific values of the mass splitting between the two HNLs: the Dirac limit corresponds to zero mass splitting, while the Majorana limit assumes a small but non-zero mass splitting that enables HNL oscillations and LNV processes~\cite{JL_benchmark}.

\begin{table}[!htbp]
\centering %
\caption{The cross-section for LNC and LNV decays, in the Dirac and Majorana limits of the 1SFH and 2QDH models. The cross-sections are shown relative to the Majorana 1SFH cross-section $\sigma_{M}$, which is given by Eq. \ref{eq:cross_section}.
The total cross-section is the sum of the LNC and LNV contributions.
}
\label{tab:xsecs:models}
\begin{tabular}{ccccc}
\toprule
& 1SFH & 1SFH & 2QDH & 2QDH \\
& Majorana limit & Dirac limit & Majorana limit & Dirac limit \\
\midrule
LNC decays & $\sigma_M$ & 2$\sigma_M$ & 2$\sigma_M$ & 4$\sigma_M$ \\
LNV decays & $\sigma_M$ &  0 & 2$\sigma_M$ & 0 \\
\midrule
Total & 2$\sigma_M$ & 2$\sigma_M$ & 4$\sigma_M$ & 4$\sigma_M$ \\
\bottomrule
\end{tabular}
\end{table}

\subsection{Background simulation}
\label{sec:bkg:simulation}
The largest SM background processes are \ttbar and \(V+\)jets production, where $V$ is a $W$ or $Z$ boson.
The \ttbar events are produced using the
\POWHEGBOX[v2]~\cite{Frixione:2007nw,Nason:2004rx,Frixione:2007vw,Alioli:2010xd}
generator at NLO with the \NNPDF[3.0nlo] PDF set
and the \hdamp parameter\footnote{The \hdamp parameter is a re-summation damping factor and one of the parameters that controls the matching of \POWHEG matrix elements to the parton shower and thus effectively regulates the high-\pT radiation against which the \ttbar system recoils.}
set to 1.5\,\mtop~\cite{ATL-PHYS-PUB-2016-020}.  The events are interfaced
to \PYTHIA[8.230] to model the parton shower,
hadronisation, and underlying event, with parameters set according
to the A14 tune~\cite{ATL-PHYS-PUB-2014-021} and using the \NNPDF[2.3lo] set of PDFs.
The decays of bottom and charm hadrons are simulated by \EVTGEN[1.6.0].

Additional \ttbar simulated samples are generated using the same \POWHEGBOX[v2] setup, but interfaced with \PYTHIA[8.230] with the parameter \texttt{pthard} set to one~\cite{10.21468/SciPostPhys.12.1.010}.
As described in Ref.~\cite{10.21468/SciPostPhys.12.1.010}, this parameter regulates the definition of the vetoed region of the showering. Its variation is used to estimate an uncertainty in the matching procedure between the matrix element (ME) generator and the parton shower (PS) algorithm, as explained in Section~\ref{sec:bkg:modelling:uncertainties}.

The production of \(V+\)jets is simulated with the \SHERPA[2.2.11]~\cite{Bothmann:2019yzt}
generator using NLO MEs for up to two partons, and LO MEs
for up to five partons calculated with the \textsc{Comix}~\cite{Gleisberg:2008fv}
and \OPENLOOPS~\cite{Buccioni:2019sur,Cascioli:2011va,Denner:2016kdg} libraries. They
are matched with the \SHERPA parton shower~\cite{Schumann:2007mg} using the \MEPSatNLO
prescription~\cite{Hoeche:2011fd,Hoeche:2012yf,Catani:2001cc,Hoeche:2009rj}
with the set of tuned parameters developed by the \SHERPA authors.
The \NNPDF[3.0nnlo] set of PDFs is used and the samples
are normalised to a next-to-next-to-leading-order (NNLO)
prediction~\cite{Anastasiou:2003ds}.

Samples of diboson final states (\(VV\)) are simulated with the \SHERPA[2.2.11] or 2.2.2 generator depending on the process, including off-shell effects and Higgs boson contributions, where appropriate.
Fully leptonic final states and semi-leptonic final states, where one boson
decays leptonically and the other hadronically, are simulated using
MEs at NLO accuracy in QCD for up to one additional parton
and at LO accuracy for up to three additional parton
emissions. Samples for the loop-induced processes \(gg \to VV\) are
simulated using LO-accurate MEs for up to one
additional parton emissions. The ME calculations are matched
and merged with the \SHERPA parton shower based on the Catani--Seymour dipole factorisation~\cite{Gleisberg:2008fv,Schumann:2007mg} using the MEPS@NLO prescription~\cite{Hoeche:2011fd,Hoeche:2012yf,Catani:2001cc,Hoeche:2009rj}.

The production of $t\bar{t}V$ events is modelled using the \MGNLO[2.3.3] generator at NLO with the \NNPDF[3.0nlo] PDF.
The events were interfaced to \PYTHIA[8.210] using the A14 tune and the \NNPDF[2.3lo] PDF set.
The decays of bottom and charm hadrons were simulated using the \EVTGEN[1.2.0] program.

The associated production of top quarks with \(W\) bosons (\(tW\)) is modelled by the \POWHEGBOX[v2]~\cite{Re:2010bp} generator at NLO in QCD using the five-flavour scheme and the \NNPDF[3.0nlo] set of PDFs.
The events were interfaced to \PYTHIA[8.230] using the A14 tune and the \NNPDF[2.3lo] set of PDFs.

Estimated yields of production of vector boson pairs, or associated production of top quarks and a vector boson ($t\bar{t}+V$, $tV$), are found to have a negligible contribution to the total background and thus are not considered further in this analysis.


%
\section{Reconstruction}
\label{sec:reco}

In the ATLAS experiment, traces from charged particles are reconstructed into tracks in two main steps.
The first, \textit{standard tracking}, is optimised for particles originating near the interaction point (IP), while the second, the LRT reconstruction, is optimised for the decay products of long-lived particles (LLPs).
All tracks are reconstructed from hits in the ID, with the track candidates found using a Kalman filter.
The tracks are then passed through an ambiguity resolution step to resolve overlaps and remove fake tracks before being re-fit using a global $\chi^2$ method. The tracks from the silicon pixel and SCT sub-detectors may then be extended into the TRT~\cite{IDTR-2022-04}.
The LRT uses the hits that are left over after the primary tracking pass and applies modified requirements, particularly loosening those on the transverse ($d_0$) and longitudinal ($z_0$) impact parameters of the track, to allow for increased reconstruction efficiency of displaced decay products.

Electron candidates are reconstructed using energy deposits in the electromagnetic calorimeter that are matched to tracks, either standard or LRT, in the ID~\cite{EGAM-2018-01}, in the range of $|\eta| < 2.47$, excluding the transition region between the barrel and endcap of the electromagnetic calorimeter of $1.37 < |\eta| < 1.52$.
The prompt electron must have transverse momentum \pt $>$ 27 \GeV and satisfy the \texttt{Medium} identification criteria, as well as the \texttt{Loose\_VarRad} isolation requirement~\cite{EGAM-2021-01}.
The displaced electrons are required to satisfy the \texttt{LLP VeryLoose} criteria, a modified version of the \texttt{VeryLoose} identification point~\cite{EGAM-2021-01} with all dependencies on number of pixel hits, track $|d_0|$, and $d_{0,\text{sig}}$ removed to improve efficiency, where $d_{0,\text{sig}} = d_0/\sigma_{d_0}$.
Additionally, displaced electrons are required not to have a \pt difference relative to the track \pt bigger than 50\% (referred to as DV track $\Leftrightarrow$ lepton match in Table~\ref{tab:preselection}).
In the case where the same cluster is matched to multiple tracks, the electron candidate with the tighter identification criteria is kept.
If both candidates satisfy the same identification requirements, the standard-track electron is kept.

Muon candidates are formed by matching reconstructed track segments in the MS to tracks in the ID~\cite{MUON-2018-03} (standard or LRT), also within $|\eta| < 2.5$ range.
The \textit{combined} track then uses information from both detector subsystems.
The prompt muon must have \pt $>$ 27 \GeV and satisfy the \texttt{Medium} identification criteria as well as the \texttt{PFlowLoose\_VarRad} isolation requirement~\cite{MUON-2018-03}.
The displaced muons are required to meet \texttt{LLP Medium} criteria, which are similar identification requirements to those of the prompt muons but with all requirements related to the ID tracks removed~\cite{MUON-2018-03}.
In the case that a single muon segment is matched to both a standard and large-radius track, the combined track with the higher quality is kept. If they have the same quality, the standard-track muon is kept.

\begin{figure}[htbp]
\centering
\subfloat[]{  \includegraphics[width=0.42\textwidth]{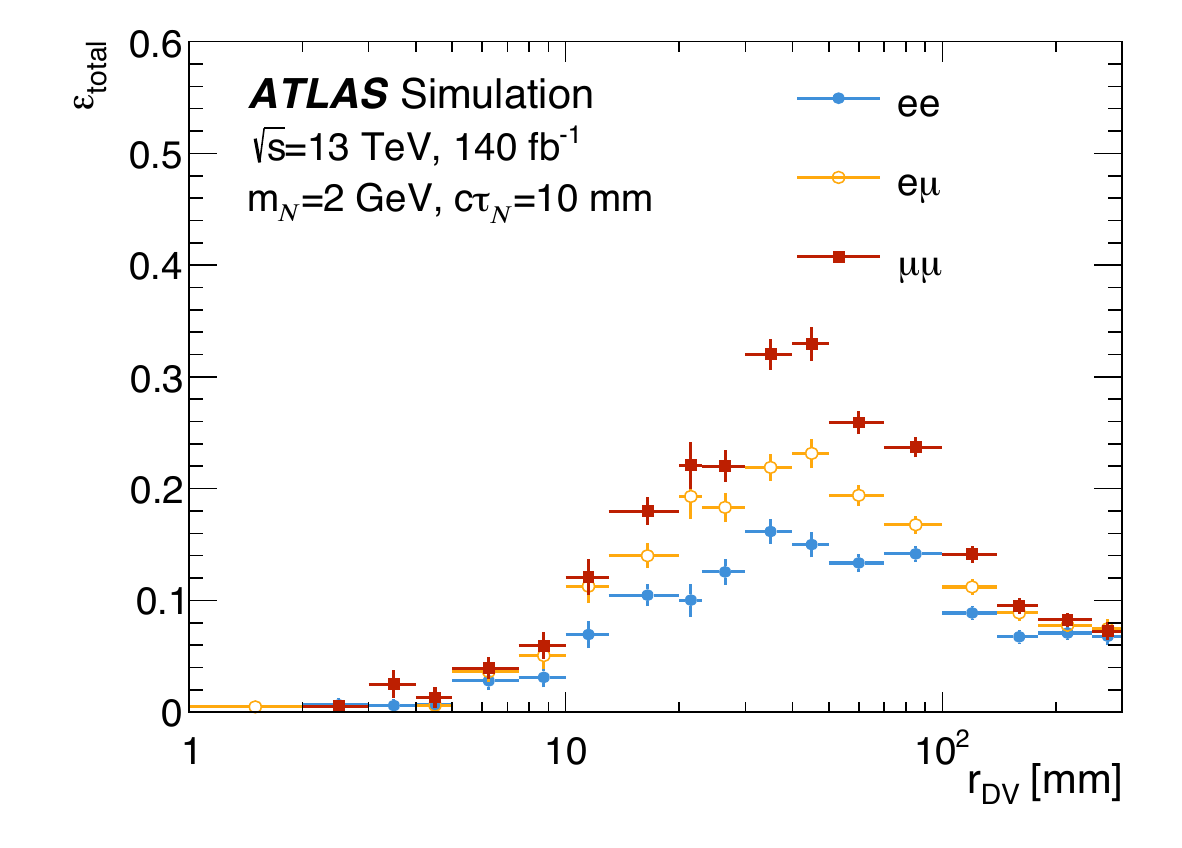}
\label{fig:reco:DVeff_ll}
}
\subfloat[]{
\includegraphics[width=0.42\textwidth]{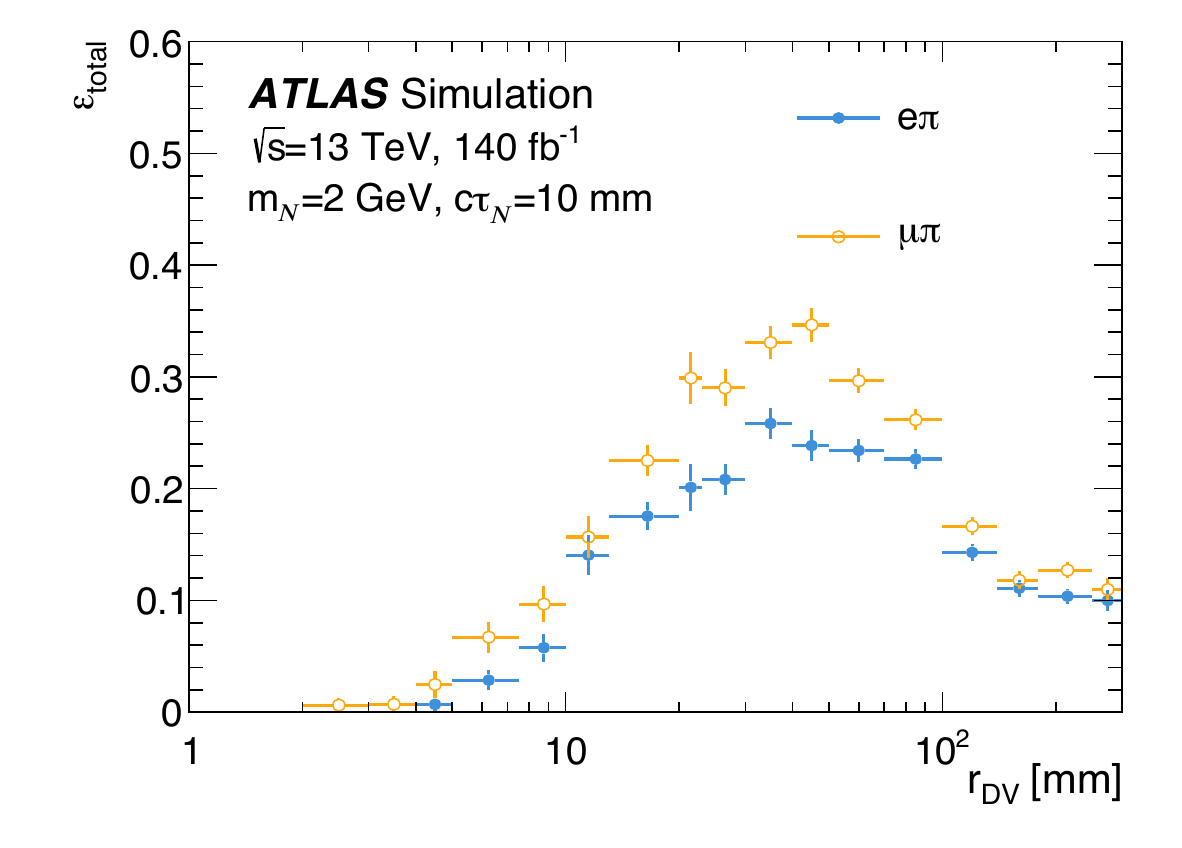}
\label{fig:reco:DVeff_lpi}
} \\
\subfloat[]{
\includegraphics[width=0.42\textwidth]{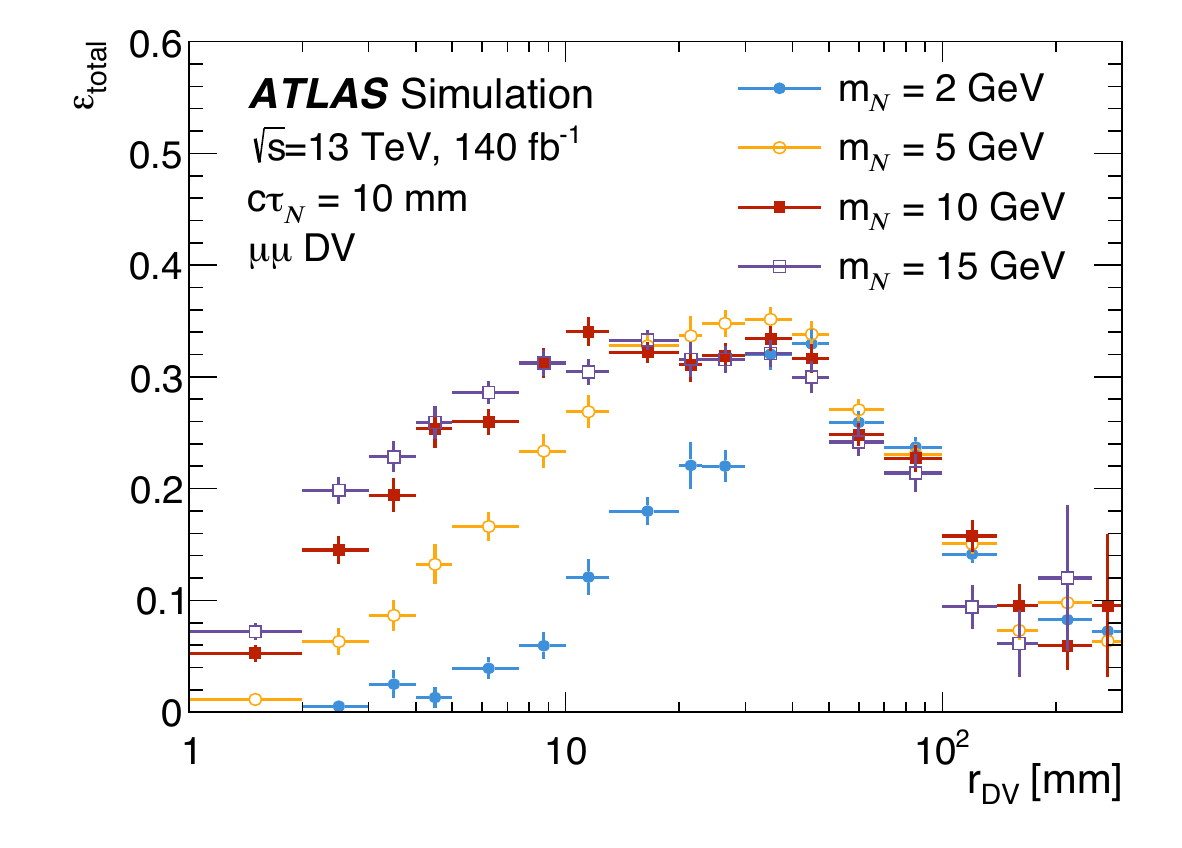}
\label{fig:reco:DVeff_uu_mass}
}
\subfloat[]{
\includegraphics[width=0.42\textwidth]{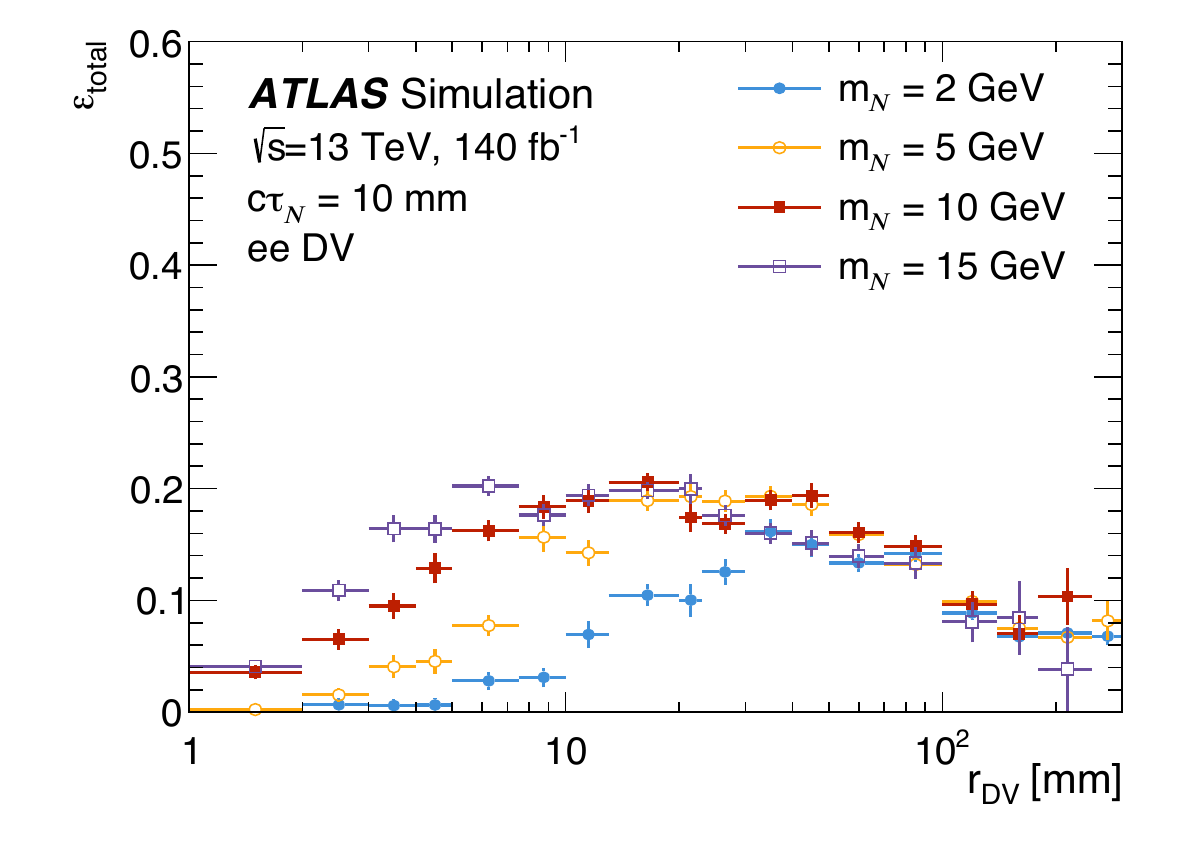}
\label{fig:reco:DVeff_ee_mass}
}
\caption{ The DV reconstruction efficiency for an HNL with $m_\myhnl$ = 2~\GeV
and $c\tau_\myhnl$ = 10 mm, as a function of the DV radius \rdv using the customised vertex reconstruction for a selection of the (a,c,d) fully leptonic and (b) semi-leptonic MC samples. Different flavour combinations are shown in (a) and (b) and decays from HNLs are shown in (c) for $\mu\mu$ DVs, and in (d) for $ee$ DVs for HNLs with various masses. }
\label{fig:reco:DVeff}
\end{figure}

The decay of the HNL is reconstructed as a displaced vertex, using tracks from both the standard and large-radius track collections.
This search uses a customised version of the ATLAS displaced vertex reconstruction~\cite{ATL-PHYS-PUB-2019-013}, designed to target both the leptonic and semi-leptonic decay modes.
The vertex reconstruction proceeds as described in Ref.~\cite{ATL-PHYS-PUB-2019-013} by first forming two-track vertices from a set of high-quality tracks.
The minimum $d_0$ requirement is loosened from $d_0> 2$~mm to $d_0> 1$~mm as a reflection of the fact that while the vertices in this search are displaced, the tracks typically point back towards the primary vertex (PV)\@.
However, when tracks are attached to the two-track vertices, unlike in Ref.~\cite{ATL-PHYS-PUB-2019-013} in which the attached tracks satisfy a very loose selection, the attached tracks in the custom vertex reconstruction must meet the same criteria as the tracks used in the initial two-track vertices.
While this search targets two-track vertices, it was found that the track attachment step decreased background contributions.
In the final selection, the vertices must contain at least one track matched to a lepton.
The efficiency to reconstruct the generated HNL decays as displaced vertices is shown as a function of the decay radius of the HNL, \rdv, in Figure~\ref{fig:reco:DVeff}, with comparable results for both leptonic and semi-leptonic vertices.
The efficiency is limited at low \rdv due to the requirement of $d_0 > 1$~mm, and at high \rdv due to decreasing number of silicon layers available, which impacts track reconstruction~\cite{IDTR-2021-03}.
The efficiency is lower for vertices with electrons due to lost momentum from bremsstrahlung and interactions with detector materials.
Lighter HNLs are more boosted, and thus likely to decay later in the detector and are more likely to fail the secondary vertex reconstruction.

Hadronic jets are reconstructed with a particle-flow algorithm~\cite{PERF-2015-09}, using a combination of information from the calorimeters and reconstructed tracks.
The jet reconstruction uses the anti-k$_t$ algorithm~\cite{Cacciari:2008gp,Fastjet} with a radius parameter $R = 0.4$.
The jets are calibrated using in situ measurements and MC simulation~\cite{JETM-2018-05}.
Jets with $\pt > 20$~\GeV and within $|\eta| < 2.5$, the $\eta$ range of the ID, are considered in this analysis.
Jets containing $b$-hadrons, known as $b$-jets, are identified using the DL1d multivariate discriminant that uses track properties, with an 85\% efficiency working point~\cite{FTAG-2018-01}. %
The $b$-tagging procedure exploits tracking and vertex information and thus is applied up to $|\eta| < 2.5$.
These jets are vetoed in the signal regions, to reduce background components.


%
\section{Trigger and event selection}
\label{sec:selection}

\label{sec:eventselection:SR}
This search requires the prompt lepton produced in the decay of the $W$ boson to be selected by the lowest-threshold unprescaled single-lepton ($e$,$\mu$) triggers from each year during Run~2, with the lepton \pt threshold varying between 20~\GeV and 26~\GeV, depending on the lepton flavour and data taking period~\cite{TRIG-2018-05,TRIG-2018-01}.

The offline event selection is divided into two stages: a pre-selection and a tighter signal-region (SR) selection.
Pre-selected events that do not satisfy the SR selection compose some of the control regions (CR) used in the background estimate.

\subsection{Pre-selection}
Events are required to have at least one PV, containing at least two standard tracks with \pt greater than 500 \MeV. If more than one such PV exists, the one with the largest $\Sigma \pt^2$ is selected, where the sum is over all tracks in the vertex.
Furthermore, events are required to include a prompt lepton matched to the lepton that fired the trigger.
It must also fulfil the track-to-vertex association (TTVA) criteria such that for electrons (muons) $|d_{0,\text{sig}}| < 5\,( < 3)$ and  $|z_{0}\sin{\theta}| < \SI{0.5}{mm}$, where $z_{0}$ is the impact parameter in the longitudinal plane.

Events are required to contain a reconstructed DV that contains exactly two tracks.
In the leptonic channel, both tracks must be matched to leptons, and in the semi-leptonic channel exactly one track must be matched to a lepton.
The DV is required to be in the fiducial volume, defined as 4 mm $< r_{\mathrm{DV}} <$ 300 mm in the leptonic channel and 20 mm $< r_{\mathrm{DV}} <$ 300 mm in the semi-leptonic channel.
The requirement is more restrictive in the semi-leptonic channel due to larger contributions from SM backgrounds.
Furthermore, DVs reconstructed in correspondence with material layers are removed via a material veto~\cite{SUSY-2018-13} that is applied in the $\ell - ee$ channel and all semi-leptonic channels to remove background due to displaced tracks that arise from interactions between primary particles and detector materials.
To reduce background from light-multijet events, the leading (subleading) track in the DV must have $\pt > 10$ \GeV ($\pt > 5$ \GeV).
The DV is also required to be isolated from all jets in the event by $\Delta R(DV,\text{jet}) > 0.4$ to mitigate the impact of MC mis-modelling in jet and $b$-jet multiplicity distributions.

Cosmic-ray muons can fake a vertex that appears to have two back-to-back tracks. To remove these, a condition is placed on the two tracks in the vertex such that $\sqrt{ (\Sigma\eta)^{2} + (\pi - \Delta\phi)^{2} } > 0.05$, in which $\Sigma\eta$ and $\Delta\phi$ consider the $\eta$ and $\phi$ of the tracks in the vertex.
To remove background from $Z$+jets production, a $Z$-boson-mass veto is applied when the prompt lepton has the same flavour as and opposite sign with respect to (one of) the displaced lepton(s): that lepton pair is required to have an invariant mass lower than 80 \GeV or greater than 100 \GeV.
The mass of the reconstructed vertex \mdv has two additional narrow mass selections, $\mdv > 0.6$~\GeV to remove \Ks decays, applied in the semi-leptonic channel; and $\mdv \notin [3.0,3.2]$~\GeV, applied to $ee$ and $\mu\mu$ vertices to remove $J/\psi$ decays in the leptonic channels.

For further suppression of DVs arising from $b$-hadron decays, DVs with $\mdv < 5$~\GeV are required to have a displacement significance $\mathcal{S} > 100$, defined as the significance of the three-dimensional distance between the DV and the PV\@.
Requiring $\mathcal{S} > 100$ generally selects DVs with a larger displacement, higher \pt, and smaller $\Delta R$ between the two tracks in the vertex.
Table \ref{tab:preselection} summarises all the pre-selection requirements.

\begin{table}[ht!]
\centering
\caption{Summary of pre-selection requirements applied}
\footnotesize\begin{tabular}{l  c | c }
\toprule
\textbf{Selection} & \textbf{Leptonic} & \textbf{Semi-leptonic} \\
\midrule
Trigger & \multicolumn{2}{c}{Lowest unprescaled single-lepton triggers, $\pt > 20 - 26$~\GeV} \\
\midrule
Prompt lepton selection & \multicolumn{2}{c}{Trigger matched electron/muon, passing \texttt{Medium} ID WP, with \pt $>$ 27 \GeV} \\
Prompt lepton TTVA & \multicolumn{2}{c}{$d_{0,\text{sig}} < 5\,(3)$ for electrons (muons), $|z_{0}\sin{\theta}| < \SI{0.5}{mm}$} \\
Prompt lepton isolation & \multicolumn{2}{c}{\texttt{Loose\_VarRad} (\texttt{PFlowLoose\_VarRad}) for electron (muon)} \\
\midrule
Displaced vertex & \multicolumn{2}{c}{Exactly 2 opposite-sign tracks, with}  \\
& $\SI{4}{mm} < r_{\text{DV}} < \SI{300}{mm}$ & $\SI{20}{mm} < r_{\text{DV}} < \SI{300}{mm}$ \\
Displaced tracks $\pt$ & \multicolumn{2}{c}{Leading (subleading) track $\pt > 10~(5)$~\GeV} \\
Displaced-leptons ID WP & \multicolumn{2}{c}{$e$: \texttt{LLP VeryLoose}, $\mu$: \texttt{LLP Medium}} \\
DV track $\Leftrightarrow$ lepton match & \multicolumn{2}{c}{Only for $e$, |$\text{track}\,\pt-\text{lepton}\,\pt| / \text{track} \pt < 0.5$} \\
$\Delta R$ selection & \multicolumn{2}{c}{$\Delta R(DV,\text{jet}) > 0.4$} \\
\midrule
Cosmic veto & \multicolumn{2}{c}{$\sqrt{ (\Sigma\eta)^{2} + (\pi - \Delta\phi)^{2} } > 0.05$}\\
$Z$-boson-mass veto & \multicolumn{2}{c}{Same-flavour opposite-sign leptons invariant mass, $m_{\ell\ell}$ not in $[80,100]$ \GeV} \\
\Ks veto & - & $\mdv > 0.6$~\GeV \\
$J/\psi$ veto & $ee$, $\mu\mu$ vertices, \mdv $\notin$ [3.0,3.2]~\GeV & -\\
Material map veto & Applied in $ee$ channels & Applied in all channels \\
\midrule
DV discriminant variable & \multicolumn{2}{c}{$\mathcal{S} > 100$ \text{if} $m_{\text{DV}} < 5$ \GeV}\\
\bottomrule
\end{tabular}
\label{tab:preselection}
\end{table}

\subsection{Signal regions}
\label{sec:selections:SR}
The SRs are defined with additional criteria to further suppress background while maintaining high signal efficiency.
A $W$-boson mass selection is placed on the invariant mass of the visible decay products of the system.
In the semi-leptonic channels, the selection requires 70 \GeV $<$ \mllpi $<$ 90 \GeV.
In the leptonic channels, due to the unaccounted for neutrino, the $W$-boson mass selection on \mlll, is looser: 40 \GeV $<$ \mlll $<$ 90 \GeV.
Additionally, isolation requirements are placed on the lepton(s) in the DV\@.
The electrons (muons) in the DV are required to satisfy the same \texttt{Loose\_VarRad} (\texttt{PFlowLoose\_VarRad}) isolation requirements as the ones applied to the prompt electrons (muons).
Finally, to further suppress backgrounds from heavy flavour SM processes, the number of $b$-jets in each event is required to be zero.

To increase the number of events in the leptonic signal regions, the channels \uuu, \uue, and \uee are merged into one $\mu-\ell\ell$ channel, and \eee, \eeu, and \euu are merged into one $e-\ell\ell$ channel.
The former merged channel probes the 1SFH($\mu$) model, while the latter probes the 1SFH($e$) model.
The combination of the two probes the multi-flavour mixing 2QDH model.
This merging is not needed in the semi-leptonic channels.

\FloatBarrier


%
%
\section{Background model}
\label{sec:background}

This section describes the background model used in the statistical analysis that performs the final background estimate, described in Section~\ref{sec:statanalysis}. There are two sources of background events: semi-leptonic decays of long-lived hadrons that contain heavy-flavour ($b$, $c$) quarks, and the production of lighter SM hadrons in jets that can lead to displaced vertices with misidentified leptons.
Background from decays of heavy-flavour hadrons is estimated from MC simulations.
The level of background that arises from hadrons that are misidentified as leptons, referred to as fake-lepton background, is not well simulated and is hence estimated with a data-driven method. Coincidental track crossings that produce displaced vertices in events with genuine prompt leptons is the main background of the previous ATLAS search~\cite{EXOT-2019-29}; however, in this work this contribution is reduced to a negligible level thanks to the improvements in the LRT reconstruction~\cite{IDTR-2021-03} and the tuning of the DV selection algorithm.

\subsection{Background from heavy-flavour hadron decays}
The decay of SM long-lived hadrons, produced in association with a prompt electron or muon, can mimic the signal.
These hadrons can decay with a significant displacement from the interaction point, and produce decay chains that
contain one or two leptons, e.g., via a $b\rightarrow c\ell^{-}\bar{\nu}$ decay that may be followed by $c\rightarrow s\ell{^+}\nu$. The key feature of this background is that the reconstructed displaced vertices satisfy $m_\textrm{DV} < 5~\GeV$, due to the masses of the long-lived $b$-hadrons.

The predictions for this background component are obtained from MC simulated samples of \ttbar and vector boson production in association with heavy-flavour jets (\(V\)+HF) production, as these processes contain both isolated leptons, and $b$- and $c$-hadrons.
The background contribution from other SM processes ($t\bar{t}+V$, $tV$, $VV$) is checked in MC simulation, and found to be negligible.
It is found that in the $\ell-\ell\ell$ channels, background events originate almost entirely from the decay chain $b\rightarrow c\ell^{-}\bar{\nu}$, $c\rightarrow s\ell{^+}\nu$, with a minor component originating from single semi-leptonic decays of $c-$ or $b-$hadrons.
In the $\ell-\ell \pi$ channels, a larger fraction of events, relative to the $\ell-\ell\ell$ channel, originates from semi-leptonic decays of $b$-hadrons.

The MC simulation is compared to data in heavy-flavour control regions (HF-CRs) designed to be statistically independent from the SRs, enriched in the HF background and with negligible signal yield. Events in the HF-CRs are selected by applying the SR selections introduced in Section~\ref{sec:selections:SR}, but requiring the presence of at least one $b$-jet and expanding the region to $40 < \mllpi < 90$~\GeV in the semi-leptonic case to increase the available statistical precision. Furthermore, to suppress signal contamination in the HF-CRs, at least one lepton matched to the reconstructed DV is required to fail the isolation requirement. These requirements allow to study the background from $b$-jets that do not get reconstructed, similar to the one that populates the SR\@.
A total of ten independent CRs are defined; one for each combination of the six $\ell-\ell\ell$ and four semi-leptonic $\ell-\ell \pi$ channels.
The HF background predictions model the data well in the HF-CRs, including the prompt-lepton \pt and $\eta$, as well as kinematic distributions related to the properties of the displaced vertices, such as the DV transverse momentum and $\eta$ distributions, and the opening angle of the tracks in the DV\@. Distributions sensitive to potential fake leptons (e.g. lepton \pt) are checked, and show a good level of agreement. Comparisons between MC simulation data of the displaced-vertex invariant mass distributions in these regions are shown in Section~\ref{sec:results}. %

\subsection{Fake-lepton background}
\label{sec:bkg:fakes}
The predictions from the MC simulations are compared with the data in the \mlll and \mllpi sidebands of the SR, in which the $W$-boson-mass selection described in Section~\ref{sec:selections:SR} is inverted.
This sideband region (SB), is orthogonal to the HF-CRs and shows that once the HF background is reduced, the MC simulations are not able to fully describe the data.
Mis-modelled features are observed for low-\pt tracks associated with the displaced vertices, as well for low prompt-lepton \pt, and in the number of reconstructed jets. The mis-modelled features are due to the presence of an additional background source of multijet production. This latter background component is more relevant for final states with electrons matched to the displaced vertices, as quantified in Section~\ref{sec:results}.
A larger contribution is expected in the $\ell-\ell \pi$ channels relative to the $\ell-\ell\ell$ ones, due to the lepton identification requirement being applied to only one of the two tracks associated with the DV.
Additionally, jets could also lead to fake prompt leptons. For this reason, this background is larger in events with prompt electrons relative to events with prompt muons, due to differences between the identification of these two physics objects.

The fake-lepton background predictions for the SRs are derived in a data-driven way using the \mlll, \mllpi sidebands, via two control regions. One is the SB defined above, which features all the selections adopted for the SR definition, except the $W$-boson mass selection that is inverted.
The other control region, referred to as the relaxed-sideband (Relaxed-SB) has the same requirements of the SB, but the isolation requirement on the DV leptons is removed. The Relaxed-SB is therefore enriched in the fake-leptons background, and has a sufficient statistical precision to obtain differential predictions for the fake-leptons background in the SRs.

Event level transfer factors (TFs), from the Relaxed-SB to the SB, are derived separately for each channel to capture potential dependencies on the final state considered. They are evaluated as the ratio of the HF background-subtracted data yields in the two regions. The HF background prediction is taken from the MC simulation, and the subtraction is done under the assumption that this component of the background is well described by the MC simulations, as checked in the HF-CRs\@. The obtained values are reported in Table~\ref{tab:TF:resbkg}.
The TF is smaller for channels with displaced electrons compared to those with displaced muons.
Because the relaxed selection requirements are much looser for electrons than for muons, a larger fraction of displaced‐electron events in these regions are misidentified fakes. Consequently, when the stricter SR requirements are applied, a small fraction of those electron events remain -- resulting in a smaller transfer factor compared to the muon channels.
The relative uncertainty obtained on each of the TF values is dominated by the statistical uncertainty of the data. This uncertainty is propagated in the statistical analysis.
The large relative uncertainty for the TF in the $\mu-\ell\ell$ region is due to the fact that the fake-leptons background is almost negligible in this channel; hence it has no effect on the results.

\begin{table}
\centering
\caption{Transfer factors (TFs) between the Relaxed-SB and the SB regions. The uncertainty is propagated from the data and MC statistical uncertainties in the yields in the regions used in the TF evaluation.}
\begin{tabular}{cccccc}
\toprule
\multicolumn{6}{c}{Transfer factor values} \\
\midrule
$\mu-\ell\ell$ & $e-\ell\ell$ & \mumupi & \emupi & \muepi & \eepi\\
\midrule
$0.009^{+0.09}_{-0.009}$ & $0.12 \pm 0.08$  & $0.74 \pm 0.13$ & $0.75 \pm 0.20$ & $0.32 \pm 0.07$ & $0.41 \pm 0.07$ \\
\bottomrule
\end{tabular}
\label{tab:TF:resbkg}
\end{table}

In the statistical analysis described in Section~\ref{sec:fitmodel}, binned distributions in the discriminating variables chosen are employed. The differential predictions for the fake-lepton background are obtained from the \mlll and \mllpi relaxed sidebands. Two templates are derived via the subtraction from the data of the HF background predictions modelled by the MC simulations: one for the left sideband, $\mlll < 40~\GeV$ ($m_{\ell\ell\pi} < 70~\GeV$), and one for the right sideband, $m_{\ell\ell\ell}$, $m_{\ell\ell\pi} > 90~\GeV$.
The templates are scaled to the full SR selection using the TF reported in Table~\ref{tab:TF:resbkg}. The SR prediction is expected to lie between the two predictions, and there is no strong motivation to prefer one to the other.
Therefore, the mean of the two is taken as the nominal pre-fit template for the fake-lepton background.

A systematic uncertainty in the shape and normalisation of this background estimate is taken as the difference of the two separate differential distributions from the mean.
Further details on the uncertainties associated with this background are provided in Section~\ref{sec:uncertainties:fakes}.
%


%
\section{Statistical analysis}
\label{sec:statanalysis}

This section describes the statistical analysis and the systematic uncertainties. The analysis is based on the profile likelihood method~\cite{Cowan:2010js}; the background estimate in the signal regions is based on a simultaneous fit to the data in the six signal regions and the ten HF-CRs. The latter are used to constrain the normalisation of the HF backgrounds in the fit.
\subsection{Fit model}
\label{sec:fitmodel}
The variable used in the profile likelihood fit for the HF-CRs is the invariant mass of the reconstructed DV, using 10 equal bins from 0 to 5~\GeV. The discriminating variable adopted in the SRs is the invariant mass of the candidate HNL, \mhnl.
In the $\ell-\ell\pi$ channels this equals to the DV mass and the same binning as that of the HF-CRs is adopted. In the leptonic channels, \mhnl is calculated taking into account the conservation of the four momentum in the $W$-boson and $\mathcal{N}$ decays, assuming negligible masses for the neutrinos and using the known mass of the $W$-boson, as described in Ref.~\cite{EXOT-2019-29}. The SRs distributions are divided into 21 bins as follows: 0.5~\GeV wide bins from 0.5 to 6~\GeV, 1~\GeV wide bins up to 15~\GeV, and one bin up to 20~\GeV. This choice is motivated by the experimental resolution of the reconstructed mass of the HNL candidate. The number of background events in each bin is treated in the statistical analysis as a random variable following a Poisson distribution. The total expected background yields in each bin is given by the sum of the background predictions.

To cope with the limited number of simulated events, the shape predictions of the HF background in the SRs are obtained by scaling the predictions derived from the MC simulation without applying the $b$-jet veto, by the efficiency of this veto.
This scale is derived from MC simulation jointly for the leptonic channels and separately for each of the four semi-leptonic channels. A dedicated systematic uncertainty, described in Section~\ref{sec:bkg:modelling:uncertainties}, is introduced to take into account possible phase-space dependencies not captured by the use of an inclusive efficiency value.

The differential predictions of the fake-leptons background in the SR for the discriminating variables used in the profile likelihood are obtained for each final state from the Relaxed-SB region, as described in Section~\ref{sec:bkg:fakes}.

Freely floating factors are introduced to allow the predictions of the HF background to be corrected using data in the HF-CRs.
One normalisation factor multiplies the expected yields of the HF background in the $\ell-\ell\ell$ channels, and a different one is used for the $\ell-\ell \pi$ regions. The choice of two separate floating normalisation factors in the profile likelihood fit is driven by the different nature of the background observed in the two types of channels.
Within the same type of final state (leptonic or semi-leptonic), the same normalisation factor is applied to all the bins and regions (SRs and CRs). %
Thanks to the use of the HF-CRs, the uncertainties associated with the background predictions are constrained by the data, reducing the overall impact on the background predictions.

\subsection{Experimental systematic uncertainties}
The effect of systematic uncertainties is taken into account via Gaussian-function-constrained nuisance parameters.
Unless otherwise specified, these are considered fully correlated across the different bins and regions included in the likelihood fit.

The uncertainty in the efficiency for the reconstruction of LRTs is derived by comparing the data and the MC simulation in $K_S^0\rightarrow\pi^+\pi^-$ events~\cite{IDTR-2021-03}. This translates into an uncertainty in the semi-leptonic signals predictions of around 1.3\%. The impact on the leptonic HNL signal yields, and on the HF background yields in the SRs is negligible.
Uncertainties in prompt tracks are derived by evaluating the reconstruction efficiency on alternative simulated samples in which the amount of passive material in the detector is varied~\cite{PERF-2015-07}. The uncertainty in the predicted semi-leptonic HNL signal yields is about 1.5\%. The effects are negligible for the background components.

Uncertainties in the electron reconstruction and identification efficiencies are computed following the method described in Ref.~\cite{EGAM-2018-01}. Overall, they translate into uncertainties in the HF background predictions in the SRs of 3\% and 1\%, for the reconstruction and identification efficiency, respectively. The same level of uncertainty is observed in the signal efficiency.
A dedicated set of uncertainties is derived for the identification criteria used for the electrons in the DVs. Their impact on the HF background predictions is negligible, while on the signal predictions, the effect is between 1.3\% and 6\%, with larger values associated with models with larger $m_{\mathcal{N}}$. Uncertainties in the electron energy scale and resolution are computed following the method described in Ref.~\cite{EGAM-2018-01}. The effect on the HF predictions in the SRs is below 1\%, while for the signal predictions in the $e-\ell\ell$ SR it can be up to 2\%. Uncertainties due to the electron isolation efficiency are between 0.6\% and 2.6\%, depending on the final state considered, for both the HF and signal predictions. Uncertainties arising from correction factors used to match the performance of the electron triggers in MC to that measured in data are about $5\%$ for both signal and background simulated samples. %

Uncertainties in the standard muon reconstruction and identification efficiencies are computed following the method described in Ref.~\cite{MUON-2018-03}.
The effect of these uncertainties in the HF background predictions is below 0.3\%, and is between 0.1\% and 0.9\% for the signal samples.
Uncertainties associated with muons reconstructed via the LRT algorithm are negligible in the HF background predictions in all the channels considered, while for the signal samples it is up to 1.1\%, with larger values for signal models with larger $m_{\mathcal{N}}$. Uncertainties in the muon momentum measurement have an effect of up to 0.5\% on the HF background predictions, and up to 1.4\% on the signal predictions, with larger effects observed in the $\mu-\ell\ell$ SR\@. The uncertainties arising from the muon trigger correction factors used to match the performance in MC to that measured in data are between 4.5\% and 4.9\% for the background predictions, and between 3.6\% and 7\% for the signal efficiency.

Systematic uncertainties in the jet energy scale and resolution~\cite{JETM-2018-05} are found to have a negligible impact on the expected background and signal yields, and are therefore not included in the profile likelihood fit.
Uncertainties in the $b$-jet selection and veto efficiencies~\cite{FTAG-2018-01} translate into uncertainties in the HF background up to 0.6\%, and up to 0.4\% in the signal predictions, in all the SRs.

The uncertainty associated with the pileup reweighting procedure is between 0.5\% and 2.3\% in the HF background predictions, depending on the channel considered. The impact on the signal predictions can be up to 9\%, with the largest effects observed in the $e-\ell\ell$ SR.

The uncertainty in the combined 2015--2018 integrated luminosity is 0.83\%~\cite{DAPR-2021-01}, obtained using the LUCID-2 detector~\cite{LUCID2} for the primary luminosity measurements, complemented by measurements using the inner detector and calorimeters.

\subsection{Monte Carlo modelling and statistical uncertainties}
\label{sec:bkg:modelling:uncertainties}
Modelling uncertainties are considered for the \ttbar and \(V\)+jets MC simulations used in the search.
For \ttbar production, the uncertainty due to initial-state radiation is estimated by
simultaneously varying the \hdamp parameter and the renormalisation (\muR) and factorisation (\muF) scales used in the generation, and using the \texttt{Var3c} up/down variants of the A14 tune
as described in Ref.~\cite{ATL-PHYS-PUB-2017-007}. This translates into an uncertainty in the total HF predictions between 0.2\% and 2.9\%. The impact of final-state radiation is evaluated by varying the renormalisation scale for emissions from the parton shower up or down by a factor of two. The uncertainties in the overall HF background predictions are between 1.5\% and 16\%.
An additional uncertainty in the modelling of the NLO matching between the ME generator and the parton shower algorithm in the \ttbar MC simulation is derived following the recommendation provided in Ref.~\cite{10.21468/SciPostPhys.12.1.010}, comparing the nominal setup with the alternative one described in Section~\ref{sec:bkg:simulation}. The resulting uncertainty in the HF background yields is between 1.4\% and 2\%.

Uncertainties from missing higher orders in the simulation of \(V\)+jets and \ttbar processes are evaluated~\cite{Bothmann:2016nao} separately by building the envelope of seven variations of the QCD
factorisation and renormalisation scales in the matrix elements, varied by factors of 0.5 and 2, while avoiding variations in opposite directions. The effect of this uncertainty in the HF background prediction is between 0.2\% and 2.1\%.

Uncertainties in the nominal PDF set are evaluated separately for both \ttbar and \(V\)+jets samples using the NNPDF replicas. The effect of the uncertainty in the strong coupling constant \(\alphas\) is assessed by variations of \(\pm 0.001\). These two sources of uncertainty are combined following the PDF4LHC recommendation~\cite{Butterworth:2015oua}. Overall, the effect on the HF background predictions is about 2.5\%.

The uncertainty associated with the scaling done to derive the SR templates for the HF predictions, described in Section~\ref{sec:fitmodel}, is 8.7\% for the \leptonicDV~channels, and between 7.4 and 15\% for the \semileptonicDV~channels. Additionally, the finite size of the simulated background samples leads to uncertainties of between 18\% and 25\% in the HF background predictions.

For the HNL signal samples, a 5\% normalisation uncertainty is assigned to account for the effect of higher-order QCD corrections to the HNL hadronic decay width~\cite{Bondarenko:2018ptm}. An additional 3\% uncertainty is included to account for the uncertainties in the cross-section and kinematics of $W$-boson production. Finally, a 5\% uncertainty is applied to the semi-leptonic HNL signal samples to account for kinematics effects unaccounted in the LO QCD simulation used.

\subsection{Fake-lepton background systematic uncertainties}
\label{sec:uncertainties:fakes}
Two sources of systematic uncertainties are considered in the profile likelihood fit for the fake-lepton background: an overall normalisation uncertainty, and a shape uncertainty. The former is derived from the uncertainty associated with the TF used to scale the templates from the Relaxed-SB to the SR selection.
The normalisation uncertainty is more than 100\% for the $\mu-\ell\ell$ channel, 65\% for $e-\ell\ell$, 18\% for $\mu-\mu\pi$, 27\% for $e-\mu\pi$, 22\% for $\mu-e\pi$ and 17\% for $e-e\pi$. The shape uncertainties in the predictions for each channel are derived as the difference between the nominal template and the two templates obtained from the low and high parts of the \mlll sidebands of the Relaxed-SB\@. The size of the uncertainty ranges between 30\% and 100\% depending on the channel and the bin considered. The nuisance parameters used for these two sources of uncertainty are fully decorrelated among the different signal regions. Additionally, bin-by-bin statistical uncertainties are taken into account via dedicated nuisance parameters.
Although the uncertainty on this background component is relatively large for the leptonic channels, the impact on the search sensitivity is limited given the size of the fake-lepton background.


%
\section{Results}
\label{sec:results}

A background-only fit to the \mhnl and \mdv distributions in the data is performed simultaneously in the SRs and in the HF-CRs.
The normalisation of the background is mostly driven by the large statistics available in the HF-CRs. In these, for the leptonic channels the HF background yields effectively increase post fit by 2 to 10\% in these HF-CRs, with the largest effect observed in the $e-\mu\mu$ HF-CR\@. For the semi-leptonic channels, the effective yields decrease by 6\% to 16\%. Similar effects are observed in the SRs, where, relative to the prefit background,
an increase in effective HF background yields of about 2\% and 16\% is observed in the $\mu-\ell\ell$ and $e-\ell\ell$ channels, respectively. For the semi-leptonic SRs, a decrease in HF background yields between 5\% and 18\% is observed. The changes are due to the effect of background modelling uncertainties, as well as the uncertainties introduced to correct the normalisation of the HF and fakes backgrounds.

\begin{figure}[tpb]
\centering
\subfloat[]{
\includegraphics[width=0.38\textwidth]{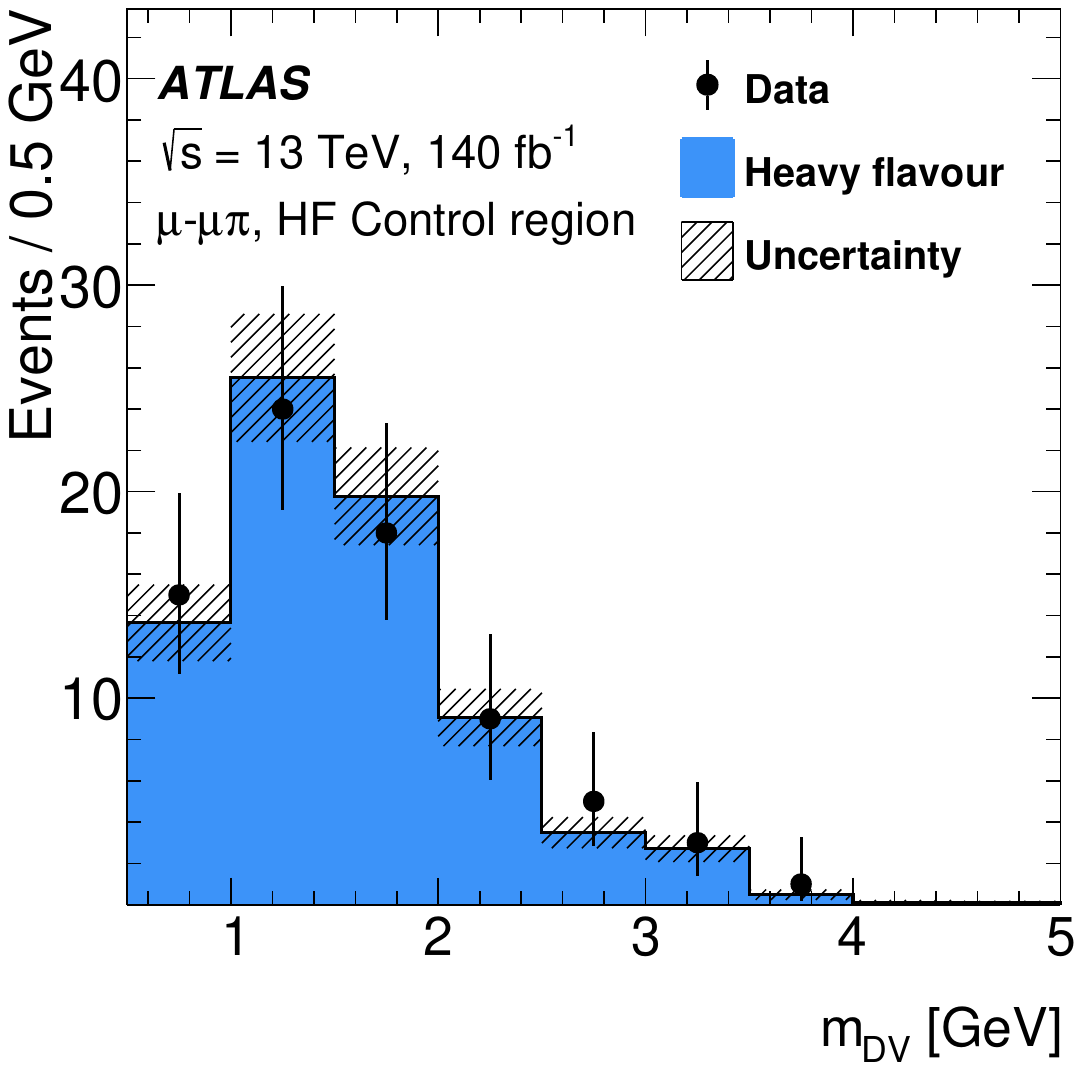}
\label{fig:datamc:CRuupi}
}
\subfloat[]{
\includegraphics[width=0.38\textwidth]{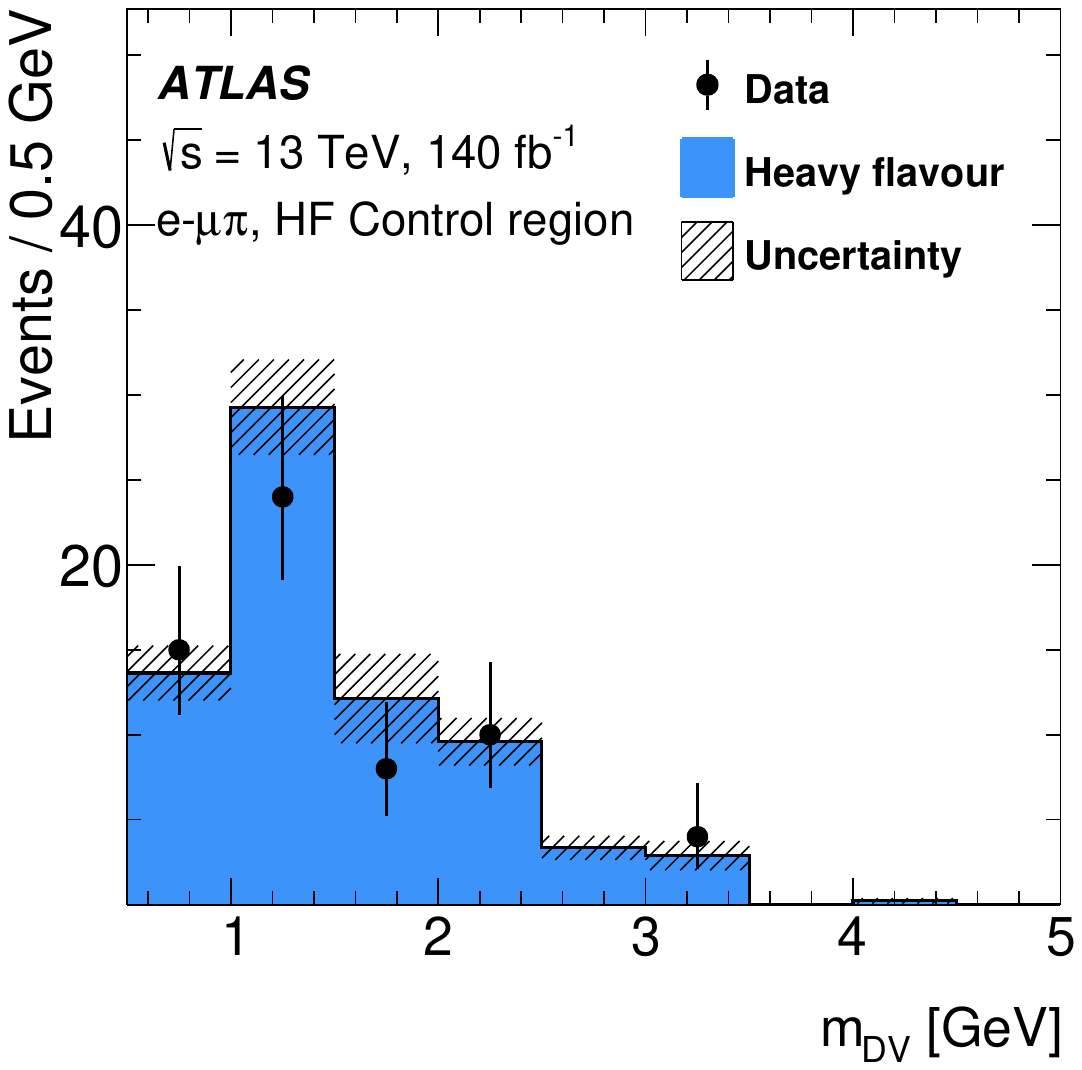}
\label{fig:datamc:CReupi}
}\\
\subfloat[]{
\includegraphics[width=0.38\textwidth]{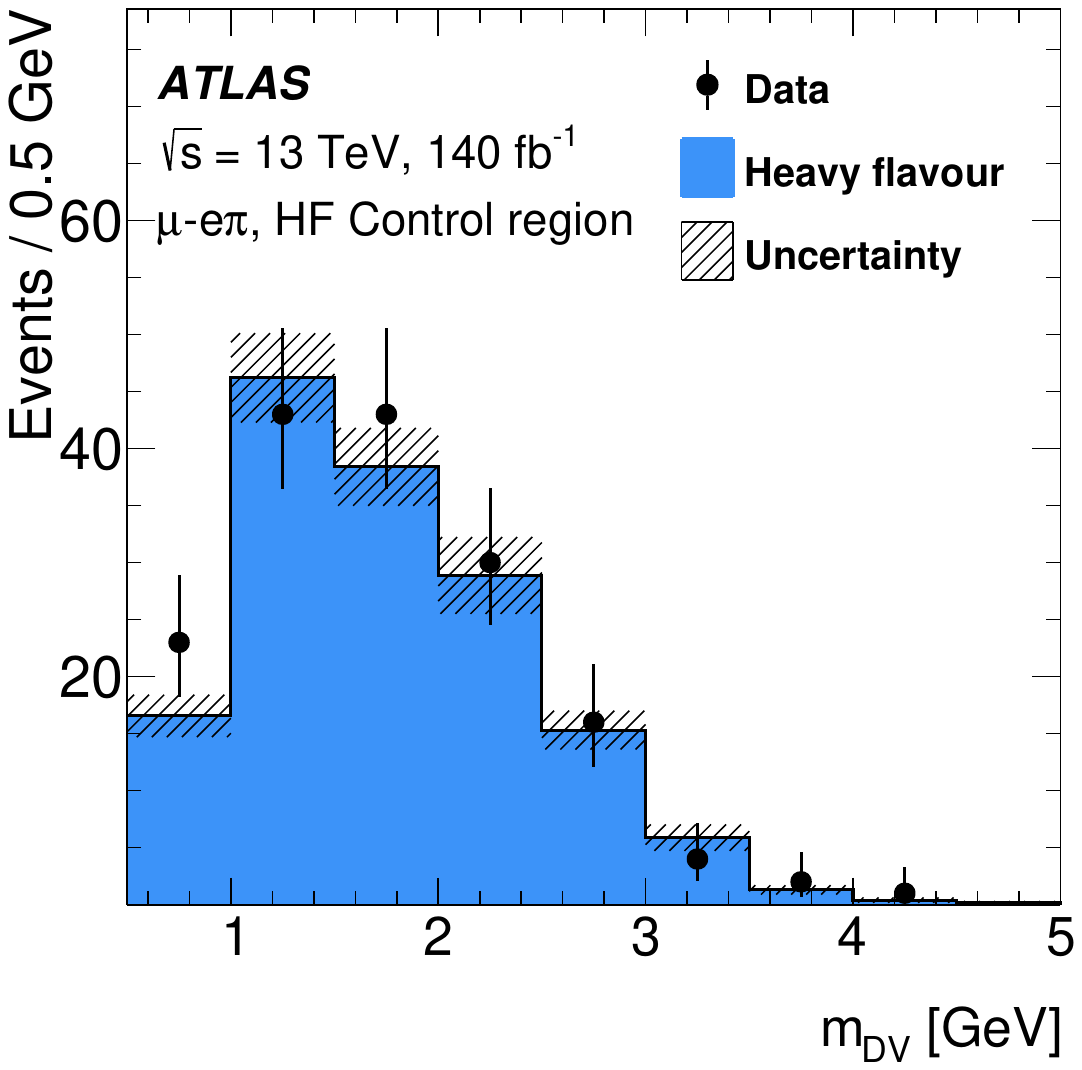}
\label{fig:datamc:CRuepi}
}
\subfloat[]{
\includegraphics[width=0.38\textwidth]{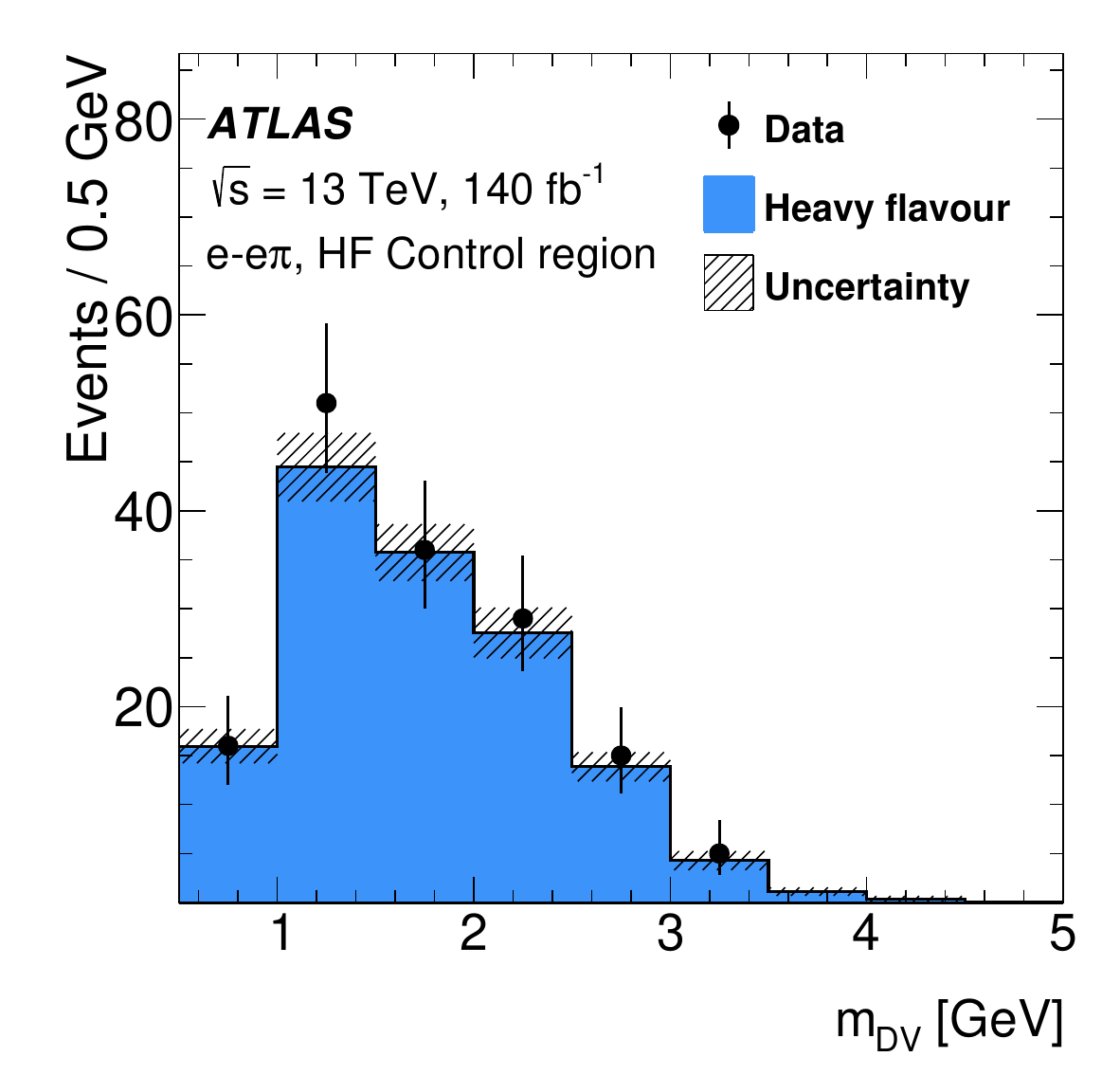}
\label{fig:datamc:CReepi}
}
\caption{Post-fit distributions in the four HF control regions for the $\ell-\ell \pi$ final states: \protect\subref{fig:datamc:CRuupi} $\mu-\mu\pi$, \protect\subref{fig:datamc:CReupi} $e-\mu\pi$, \protect\subref{fig:datamc:CRuepi} $\mu-e\pi$, \protect\subref{fig:datamc:CReepi} $e-e\pi$.
The error bands include the statistical and systematic uncertainties in the background predictions.
}
\label{fig:CR:semileptonic}
\end{figure}

\begin{figure}[htbp]
\centering
\subfloat[]{
\includegraphics[width=0.38\textwidth]{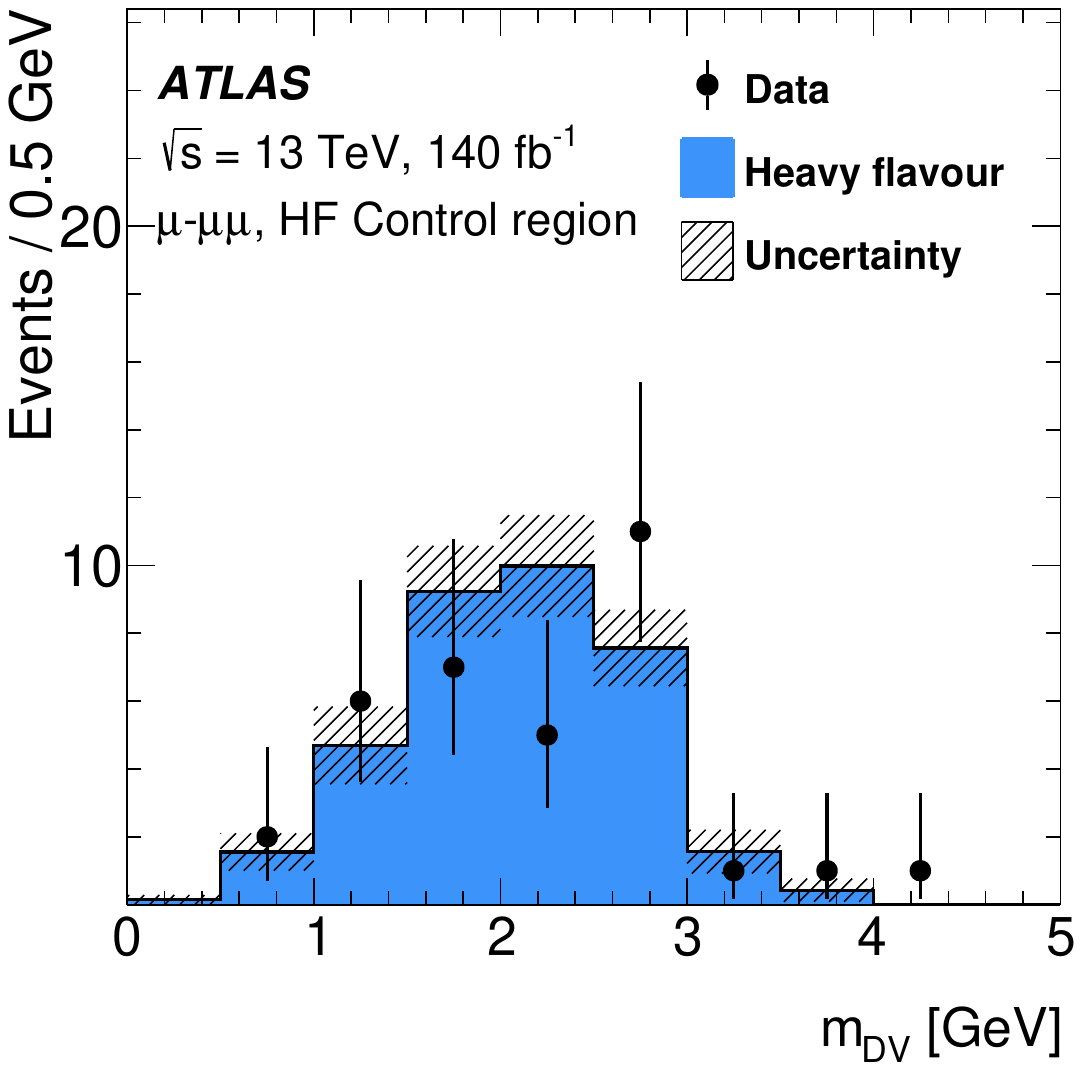}
\label{fig:datamc:CRuuu}
}
\subfloat[]{
\includegraphics[width=0.38\textwidth]{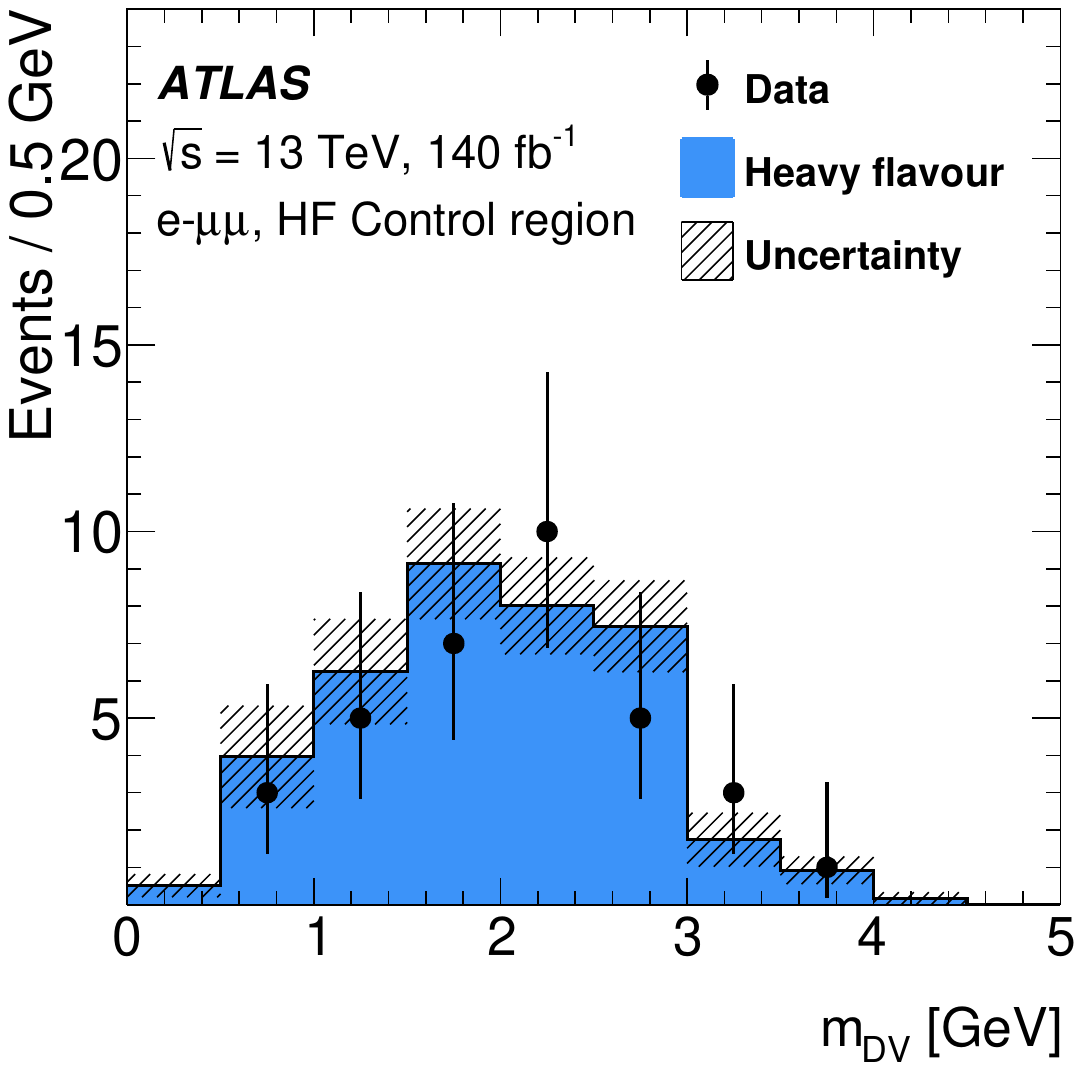}
\label{fig:datamc:CReuu}
}\\
\subfloat[]{
\includegraphics[width=0.38\textwidth]{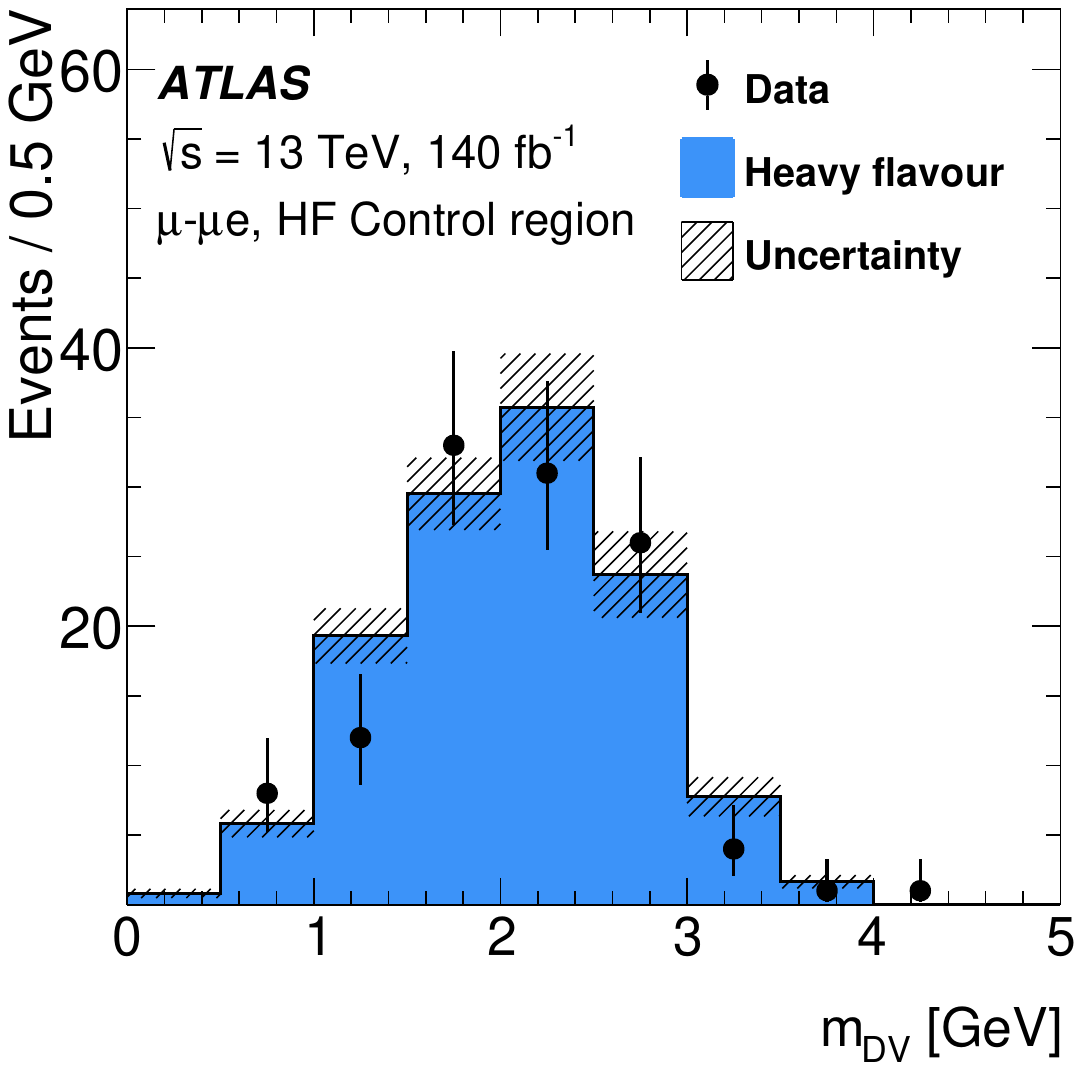}
\label{fig:datamc:CRuue}
}
\subfloat[]{
\includegraphics[width=0.38\textwidth]{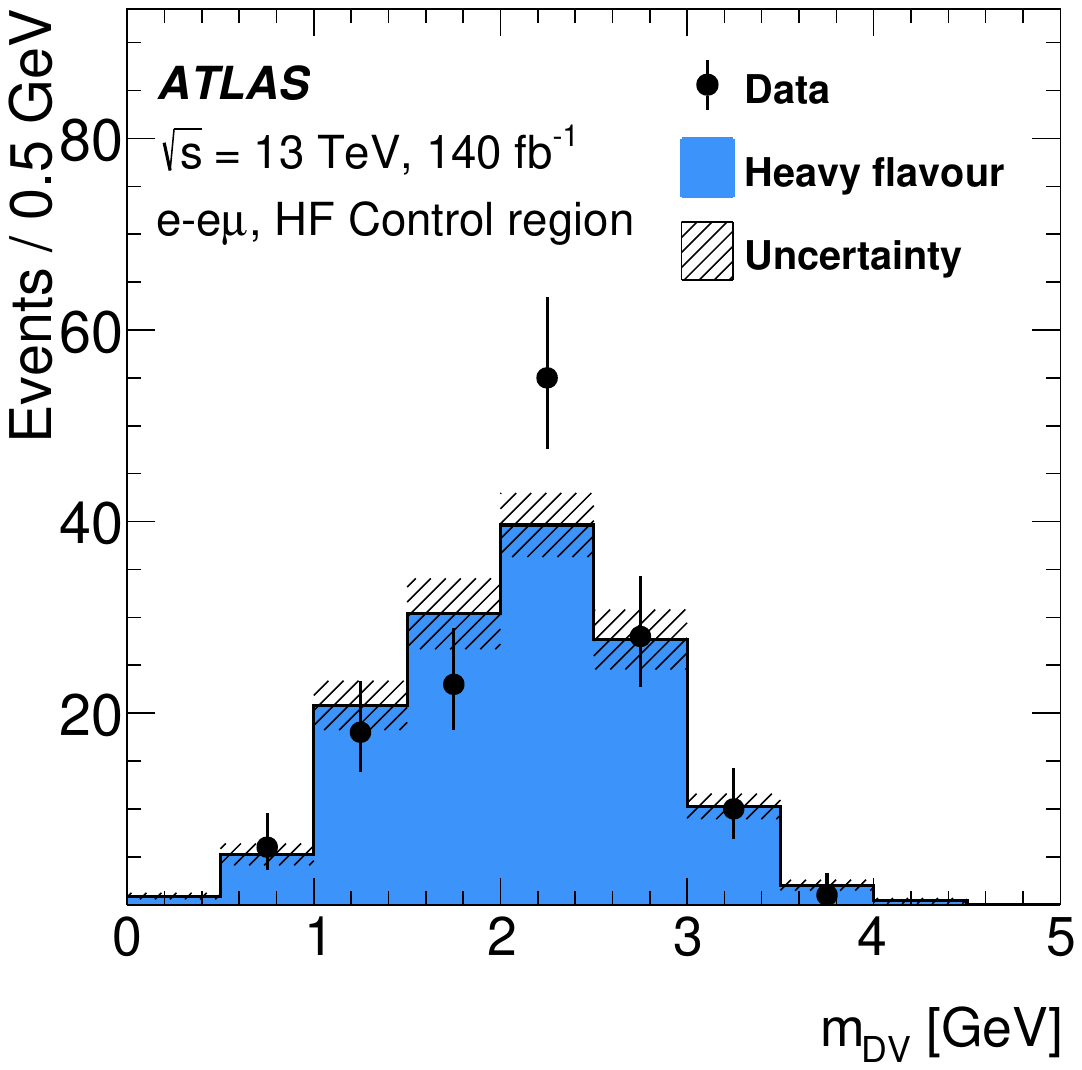}
\label{fig:datamc:CReeu}
}\\
\subfloat[]{
\includegraphics[width=0.38\textwidth]{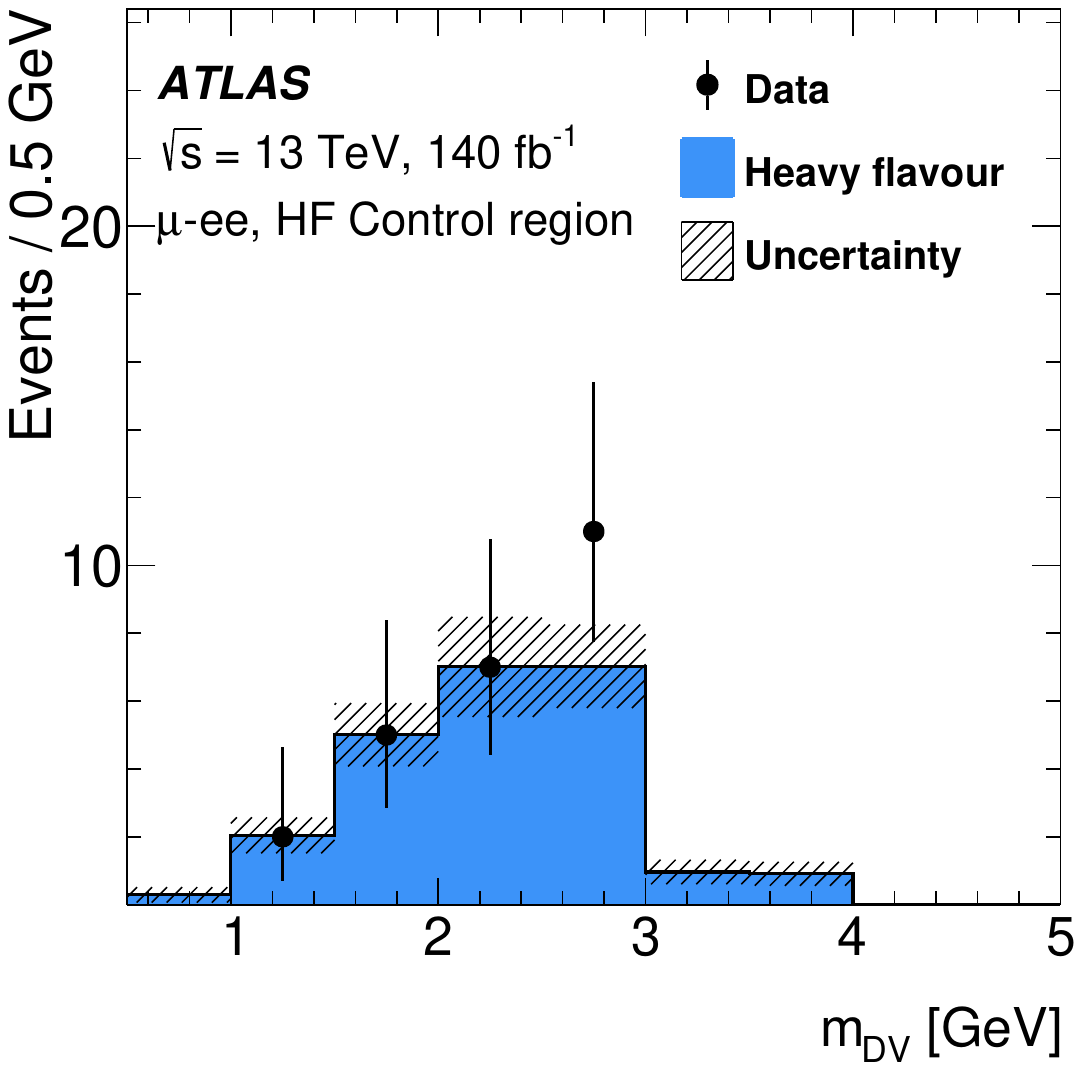}
\label{fig:datamc:CRuee}
}
\subfloat[]{
\includegraphics[width=0.38\textwidth]{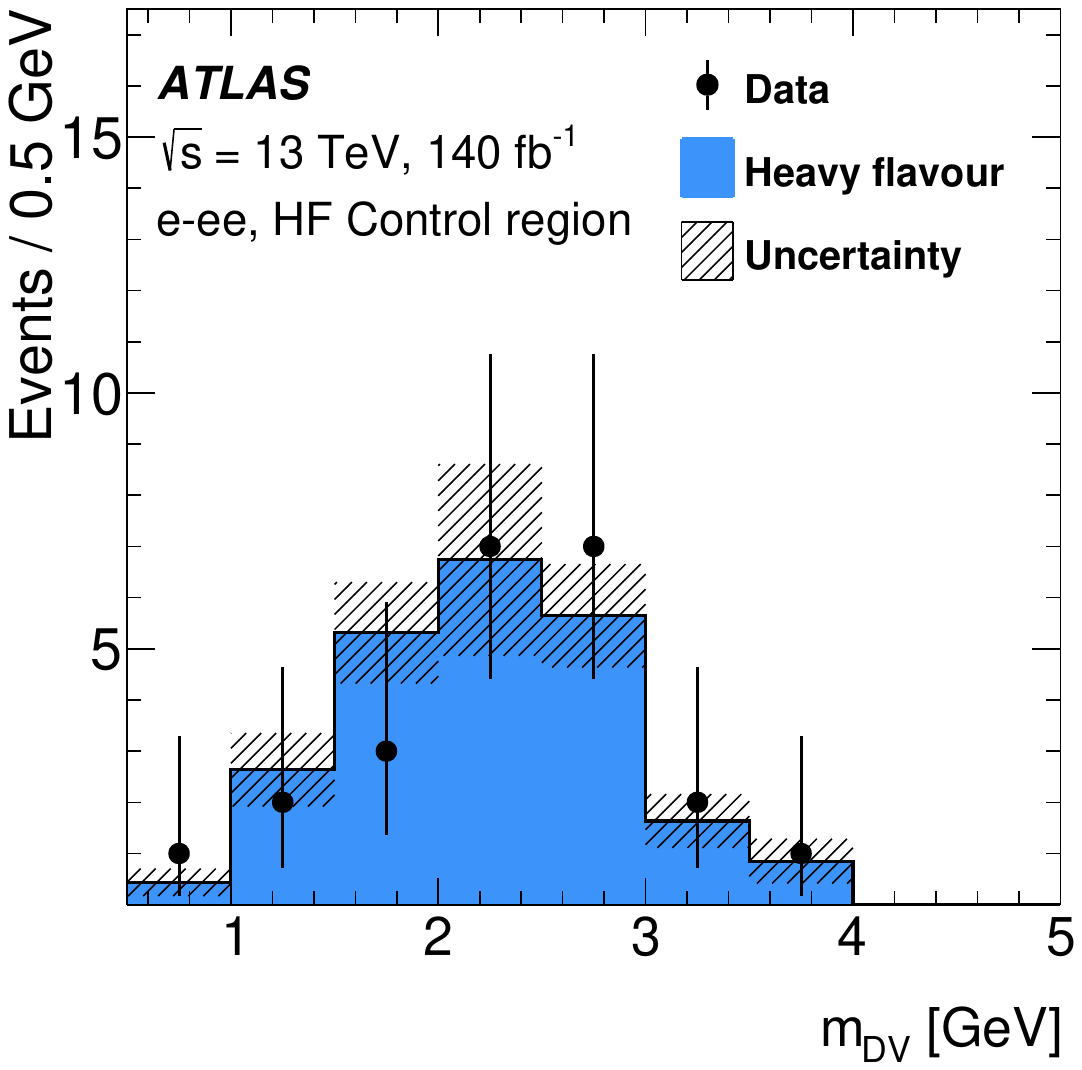}
\label{fig:datamc:CReee}
}
\caption{Post-fit distributions in the six HF control regions for the $\ell-\ell\ell$ final states: \protect\subref{fig:datamc:CRuuu} $\mu-\mu\mu$,
\protect\subref{fig:datamc:CReuu} $e-\mu\mu$,
\protect\subref{fig:datamc:CRuue} $\mu-\mu e$,
\protect\subref{fig:datamc:CReeu} $e-e\mu$,
\protect\subref{fig:datamc:CRuee} $\mu-ee$,
\protect\subref{fig:datamc:CReee} $e-ee$.
The error bands include the statistical and systematic uncertainties in the background predictions.
}
\label{fig:CR:leptonic}
\end{figure}

\begin{figure}[htbp]
\centering
\subfloat[]{
\includegraphics[width=0.35\textwidth]{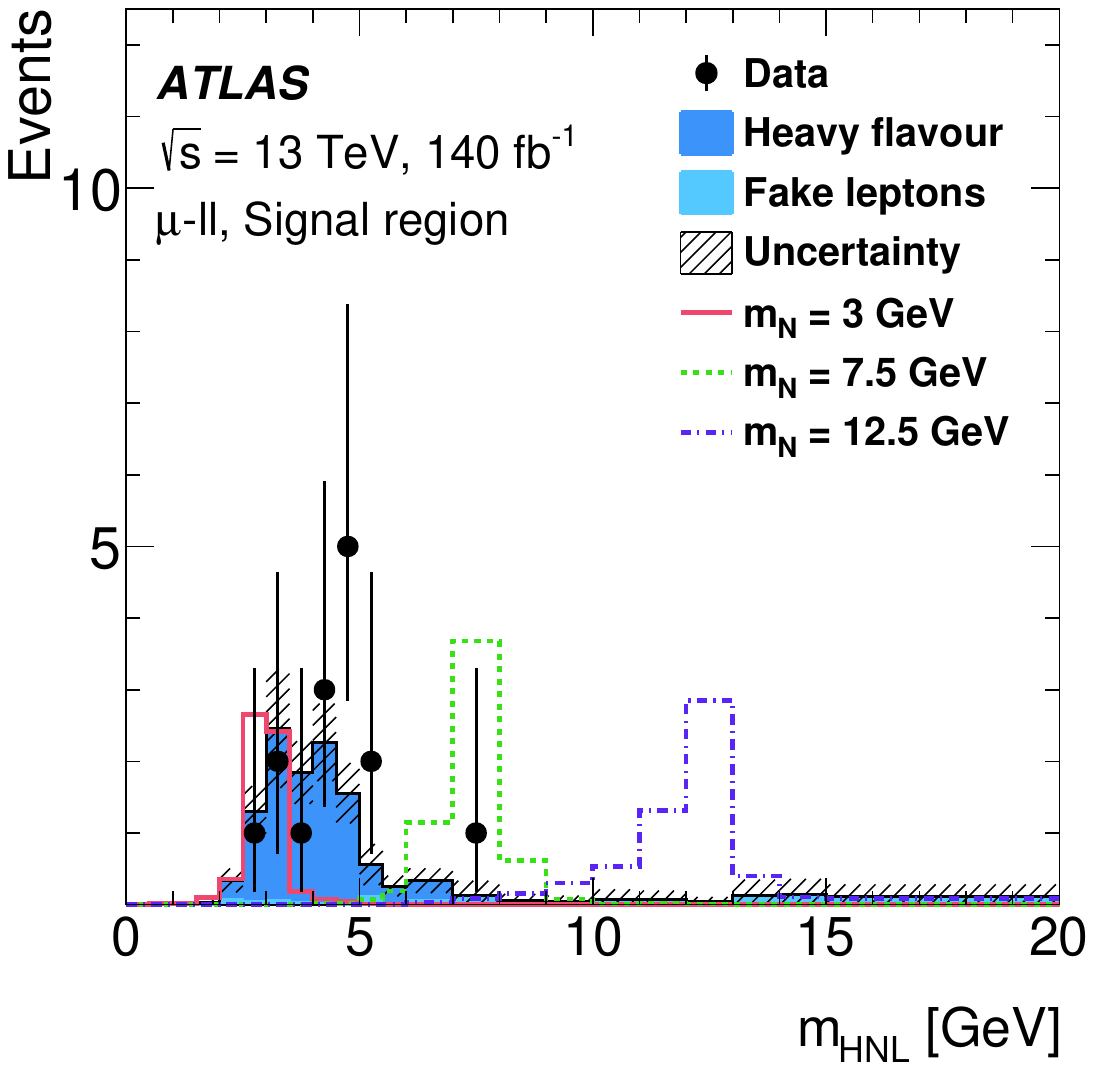}
\label{fig:datamc:SRuxx}
}
\subfloat[]{
\includegraphics[width=0.35\textwidth]{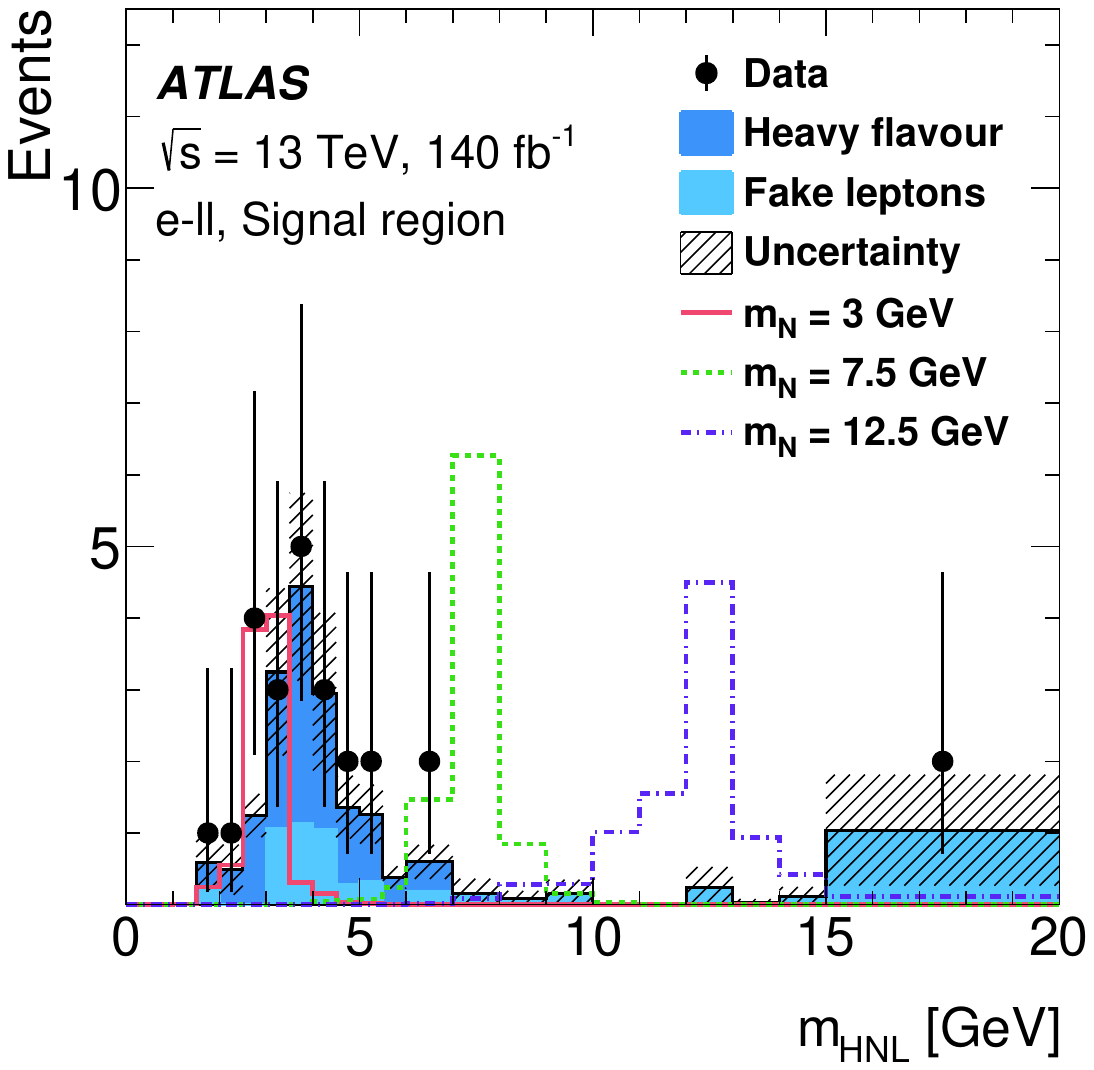}
\label{fig:datamc:SRexx}
}\\
\subfloat[]{
\includegraphics[width=0.35\textwidth]{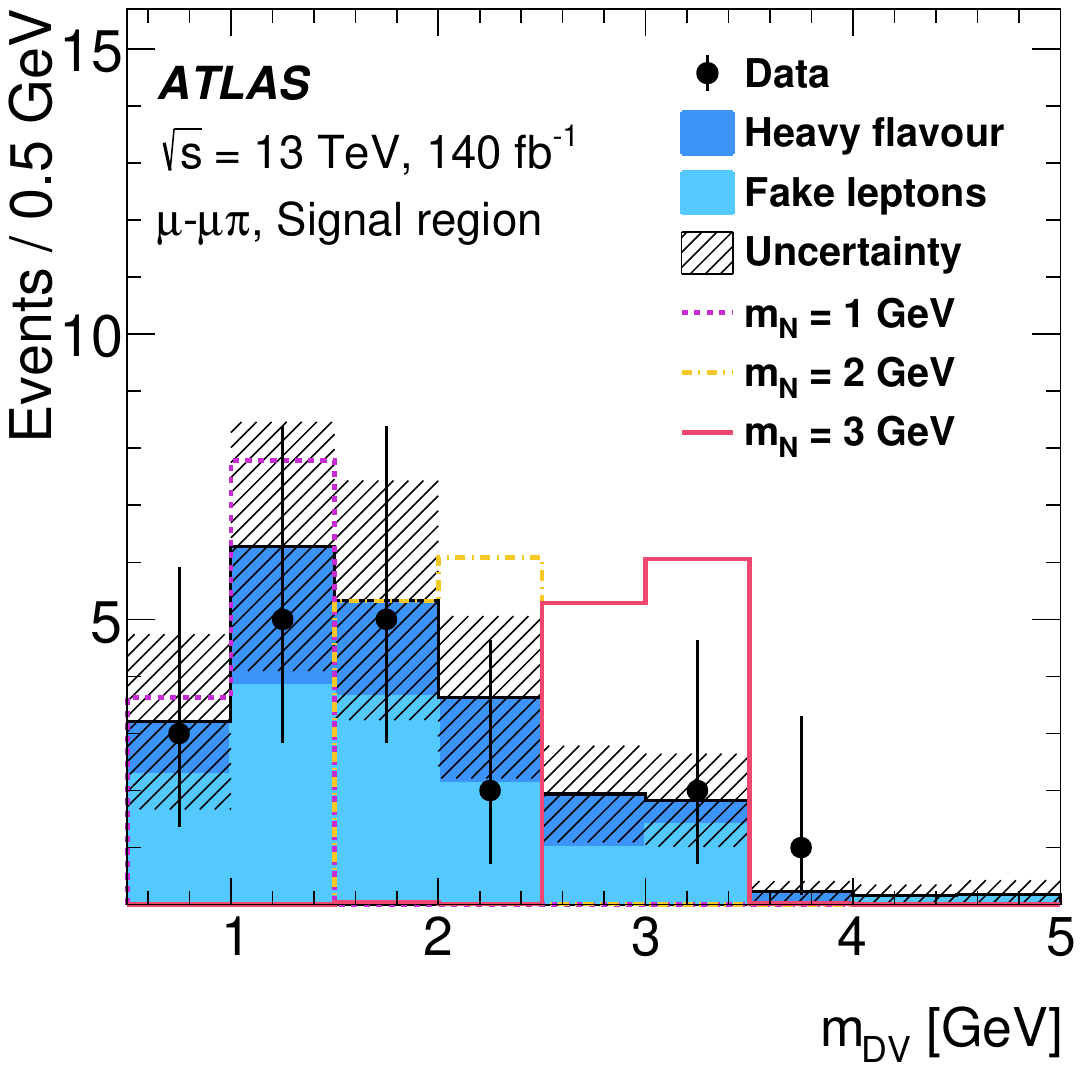}
\label{fig:datamc:SRuupi}
}
\subfloat[]{
\includegraphics[width=0.35\textwidth]{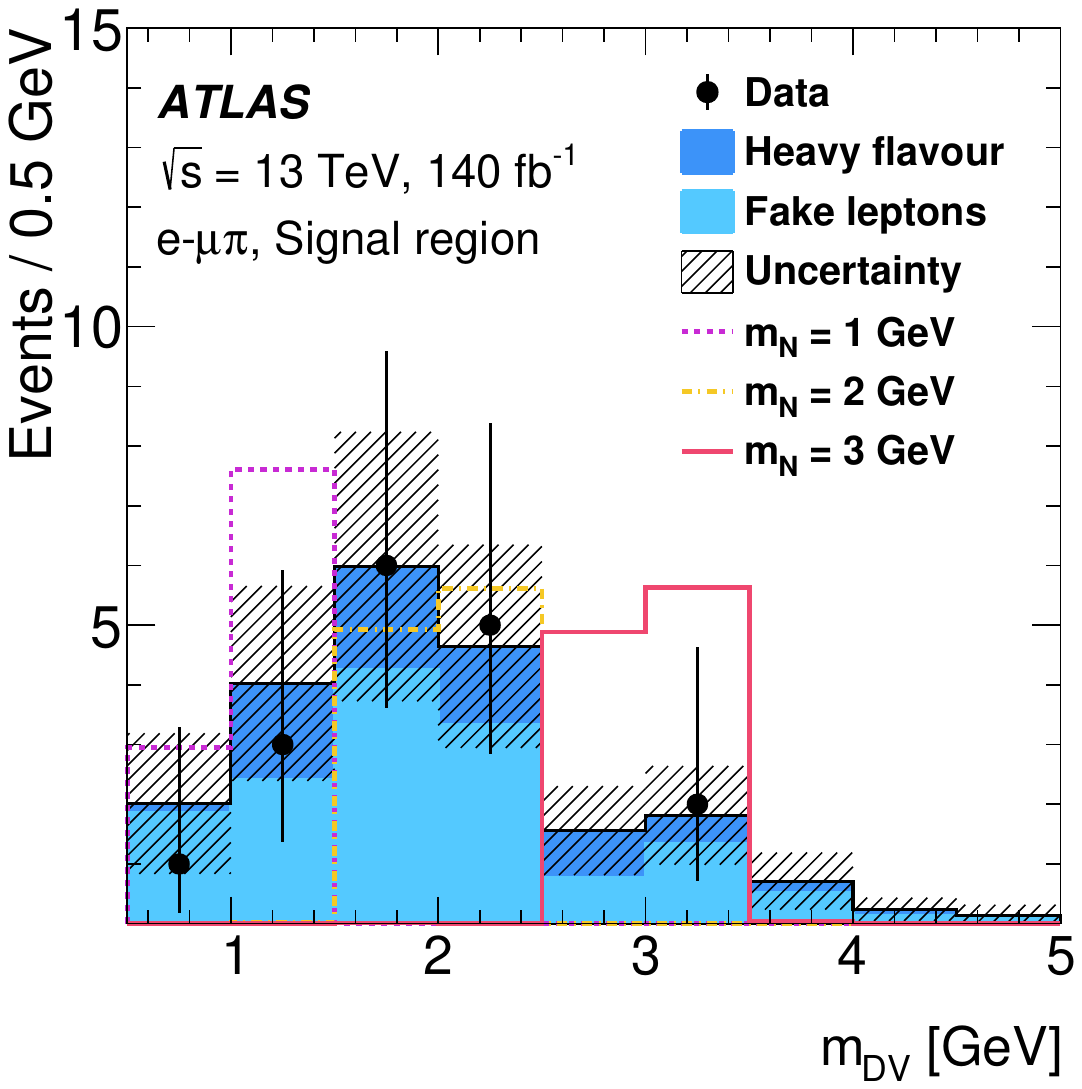}
\label{fig:datamc:SReupi}
}\\
\subfloat[]{
\includegraphics[width=0.35\textwidth]{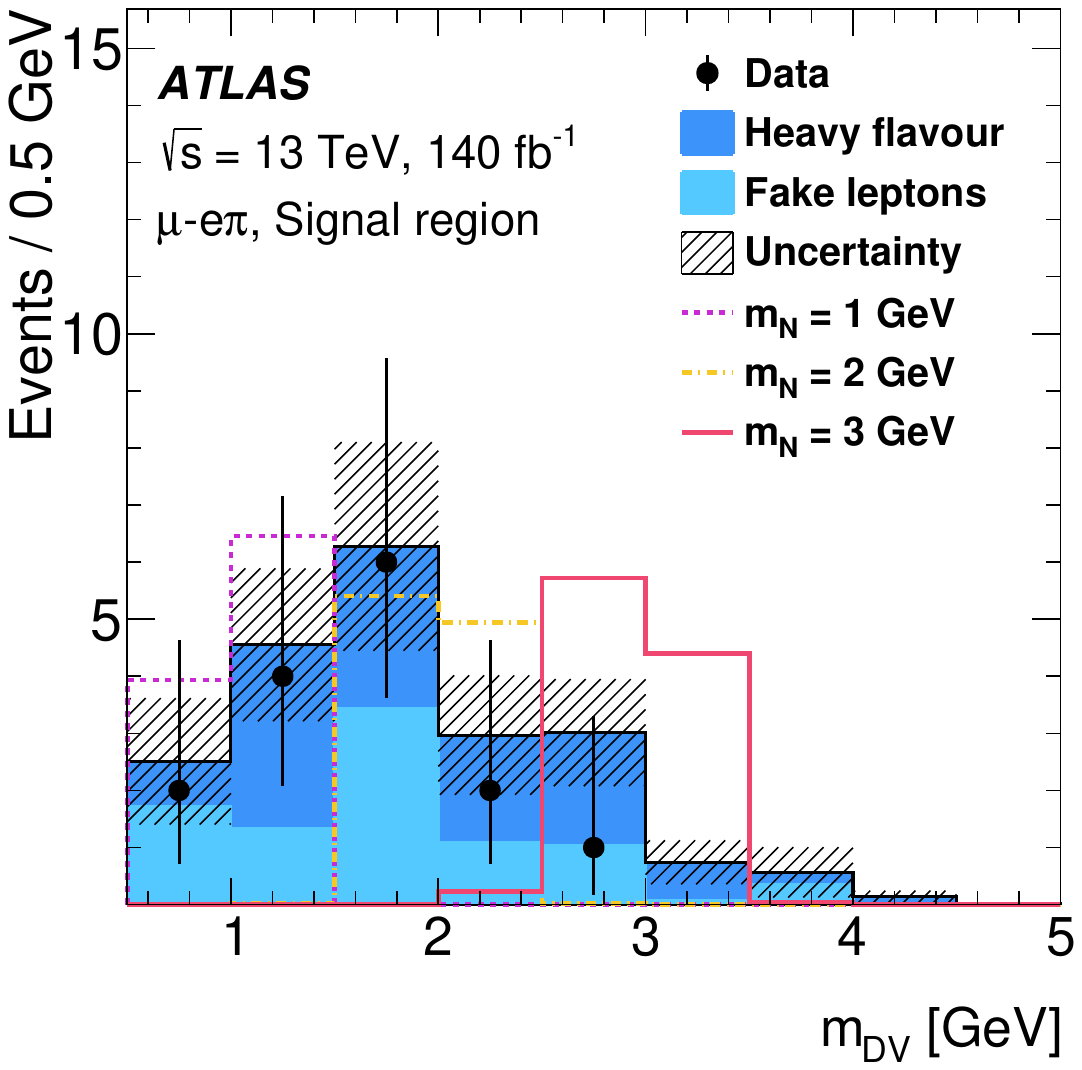}
\label{fig:datamc:SRuepi}
}
\subfloat[]{
\includegraphics[width=0.35\textwidth]{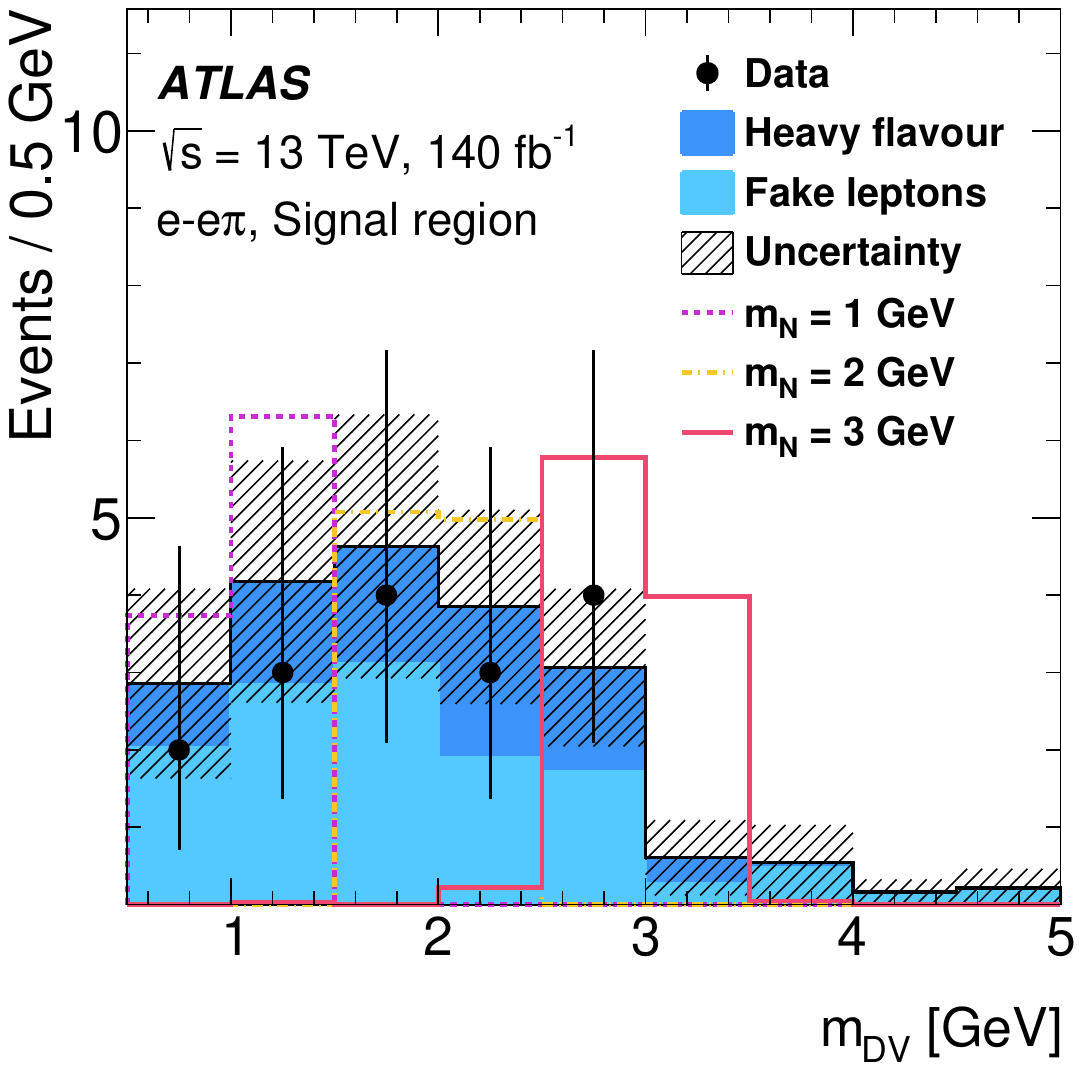}
\label{fig:datamc:SReepi}
}
\caption{Post-fit distributions in the six SRs,
\protect\subref{fig:datamc:SRuxx} $\mu-\ell\ell$,
\protect\subref{fig:datamc:SRexx} $e-\ell\ell$,
\protect\subref{fig:datamc:SRuupi} $\mu-\mu \pi$,
\protect\subref{fig:datamc:SReupi} $e-\mu \pi$,
\protect\subref{fig:datamc:SRuepi} $\mu-e \pi$,
\protect\subref{fig:datamc:SReepi} $e-e \pi$.
The error bands include the statistical and systematic uncertainties in the background predictions. The expected distribution for three example signal samples, %
normalised to half of the total background, are overlaid for illustration purposes.     }
\label{fig:SR}
\end{figure}

Figures \ref{fig:CR:semileptonic} and \ref{fig:CR:leptonic} present the comparison of \mdv between the data and background predictions in the HF-CRs of the semi-leptonic and leptonic final states, respectively. A good agreement, both in the normalisation and in the shape, is observed.
Figure \ref{fig:SR} shows the comparison for \mhnl and \mdv in the six SRs.
No significant excess over the background predictions consistent with an HNL signal is observed.
The largest local significance observed is 3.1$\sigma$, at $m_\myhnl$ of 5~\GeV in the 2QDH model, due to the upward statistical fluctuation in the 4.5--5~\GeV bin in the $\mu-\ell\ell$ SR.
The corresponding global significance is 2.0$\sigma$.

The inclusive numbers of background events predicted in the SRs are compared with the observed event yields in Table \ref{tab:SR:lep} for the \leptonicDV channels and in Table \ref{tab:SR:semilep} for the \semileptonicDV channels after a background-only fit to the SRs and HF-CRs.
\begin{table}
\centering
\caption{Inclusive predicted background events in the two SRs for leptonic DV final states after a background-only fit to the SRs and HF-CRs, compared with the number of observed events.}
\begin{tabular}{lcc}
\toprule
Signal region & \uxx & \exx \\
\midrule
Heavy flavour  & 10.5~$\pm$~1.5 & 12.3~$\pm$~1.7\\
Fakes  & 1.2~$\pm$~1.7 & 6.1~$\pm$~3.3\\
\midrule
Total bkg  & 11.8~$\pm$~2.2 & 18.4~$\pm$~3.3\\
\midrule
Observed data  & 15 & 25\\
\bottomrule
\end{tabular}
\label{tab:SR:lep}
\end{table}

\begin{table}
\centering
\caption{Inclusive predicted background events in the four SRs for semi-leptonic DV final states after a background-only fit to the SRs and HF-CRs, compared with the number of observed events.}
\begin{tabular}{lcccc}
\toprule
Signal region & $\mu-\mu\pi$ & $\mu-e\pi$ & $e-\mu\pi$ & $e-e\pi$ \\
\midrule
Heavy flavour  & 8.2~$\pm$~1.3 & 11.7~$\pm$~1.5 & 6.3~$\pm$~1.3 & 7.4~$\pm$~1.1\\
Fakes  & 15~$\pm$~4 & 9.1~$\pm$~3.2 & 15~$\pm$~4 & 12.7~$\pm$~3.3\\
\midrule
Total bkg  & 23~$\pm$~4 & 20.8~$\pm$~3.3 & 21~$\pm$~4 & 20.1~$\pm$~3.3\\
\midrule
Observed data  & 18 & 15 & 17 & 16\\
\bottomrule
\end{tabular}
\label{tab:SR:semilep}
\end{table}

A breakdown of the sources of uncertainty in the HF background predictions in each SR is reported in Table \ref{tab:hf:unc}. The largest single source of uncertainty in the post-fit HF background yields is associated with the overall normalisation of the HF background in the fit (HF floating normalisation). Background modelling uncertainties are larger than experimental uncertainties, whose effects are sub-dominant. Another significant source of uncertainty is the available MC statistical precision in the procedure used to obtain SR templates (SR template building).

\begin{table}[!tb]
\centering
\caption{Grouped effects of different sources of systematic uncertainties in the HF background predictions in the six SRs after a background-only fit to the SRs and HF-CRs. The different sources of uncertainty are correlated after the minimisation of the likelihood in the simultaneous SR+CR fit; therefore their sum in quadrature does not match the total uncertainty.}
\scalebox{0.9}{
\begin{tabular}{lcccccc}
\toprule
Systematic & SR $\mu-\ell\ell$ & SR $e-\ell\ell$ & SR $\mu-\mu\pi$ & SR $\mu-e\pi$ & SR $e-\mu\pi$ & SR $e-e\pi$ \\
\noalign{\smallskip}\hline\noalign{\smallskip}
Electrons & 0.2~\% & 6~\% & -- & 0.8~\% & 7~\% & 6~\% \\
Muons & 5~\% & 2~\% & 5~\% & 4~\% & 1~\% & -- \\
Flavour tagging & 0.5~\% & 0.7~\% & 0.2~\% & 0.3~\% & 0.6~\% & 0.2~\% \\
Pileup reweighting & 2~\% & 2~\% & 2~\% & 0.5~\% & 0.2~\% & 1~\% \\
\midrule
Background modelling & 12~\% & 10~\% & 10~\% & 9~\% & 13~\% & 14~\% \\
SR template building & 8~\% & 8~\% & 7~\% & 9~\% & 15~\% & 10~\% \\
MC statistics & 1.3~\% & 1.3~\% & 0.5~\% & 0.1~\% & 0.9~\% & 0.3~\% \\
\midrule
HF floating normalisation & 13~\% & 13~\% & 13~\% & 13~\% & 13~\% & 13~\% \\
\noalign{\smallskip}\hline\noalign{\smallskip}
Total & 14~\% & 14~\% & 16~\% & 13~\% & 20~\% & 15~\% \\
\bottomrule
\end{tabular}
}
\label{tab:hf:unc}
\end{table}

\FloatBarrier

\subsection{Interpretations}
Exclusion limits at 95\% CL are set on the two-dimensional plane of the $|U_{\alpha}|^2$ and $m_{\mathcal{N}}$ parameters for several HNL models, by performing a simultaneous fit to all the signal and control regions. Since the signal contamination in the CRs is negligible, the presence of signal is considered only in the SRs.
For $m_\mathcal{N}$ hypotheses above 3~\GeV, only the $\ell-\ell\ell$ signal and control regions are used, as the branching ratio for $\mathcal{N}\rightarrow\ell\pi$ drops substantially~\cite{Bondarenko:2018ptm}.

\begin{figure}[htbp]
\centering
\subfloat[]{
\includegraphics[width=0.45\textwidth]{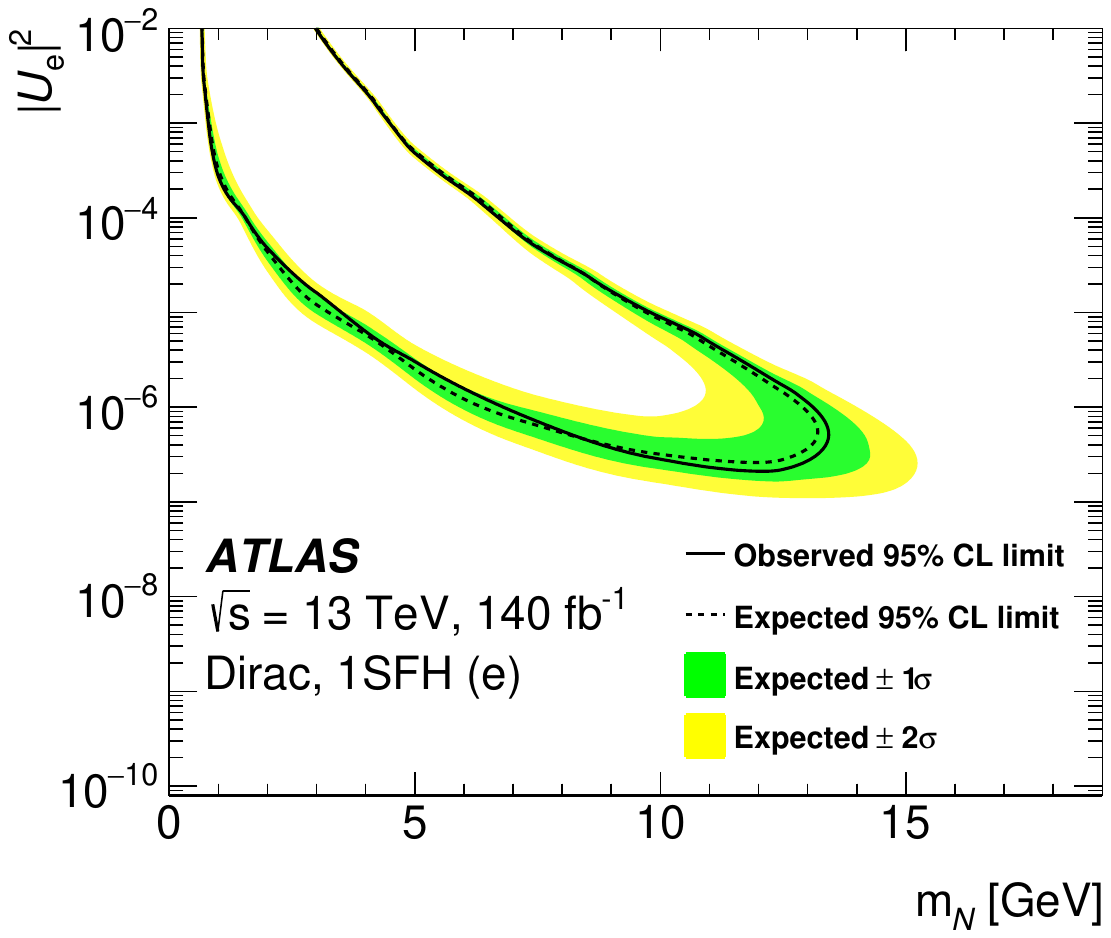}
\label{fig:limits:D_SF_EL}
}
\subfloat[]{
\includegraphics[width=0.45\textwidth]{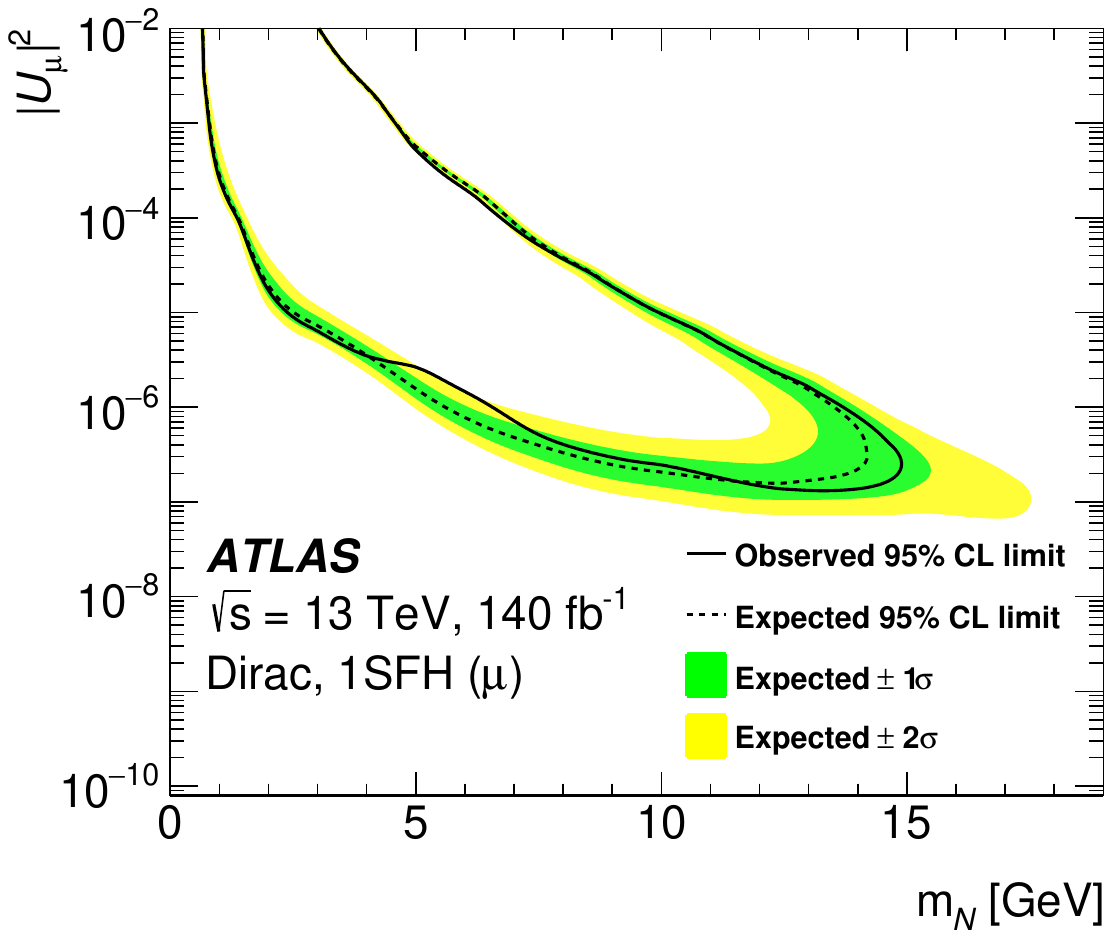}
\label{fig:limits:D_SF_MU}
}\\
\subfloat[]{
\includegraphics[width=0.45\textwidth]{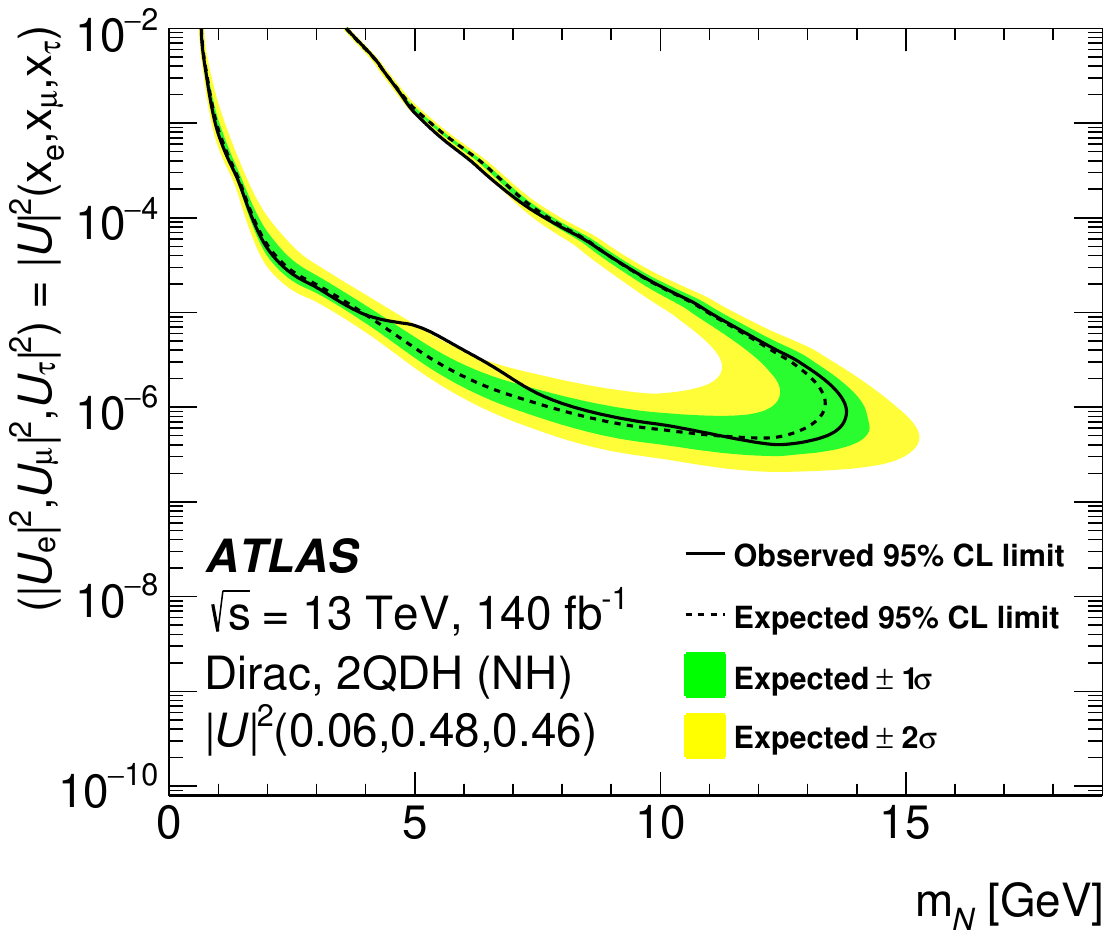}
\label{fig:limits:D_NH}
}
\subfloat[]{
\includegraphics[width=0.45\textwidth]{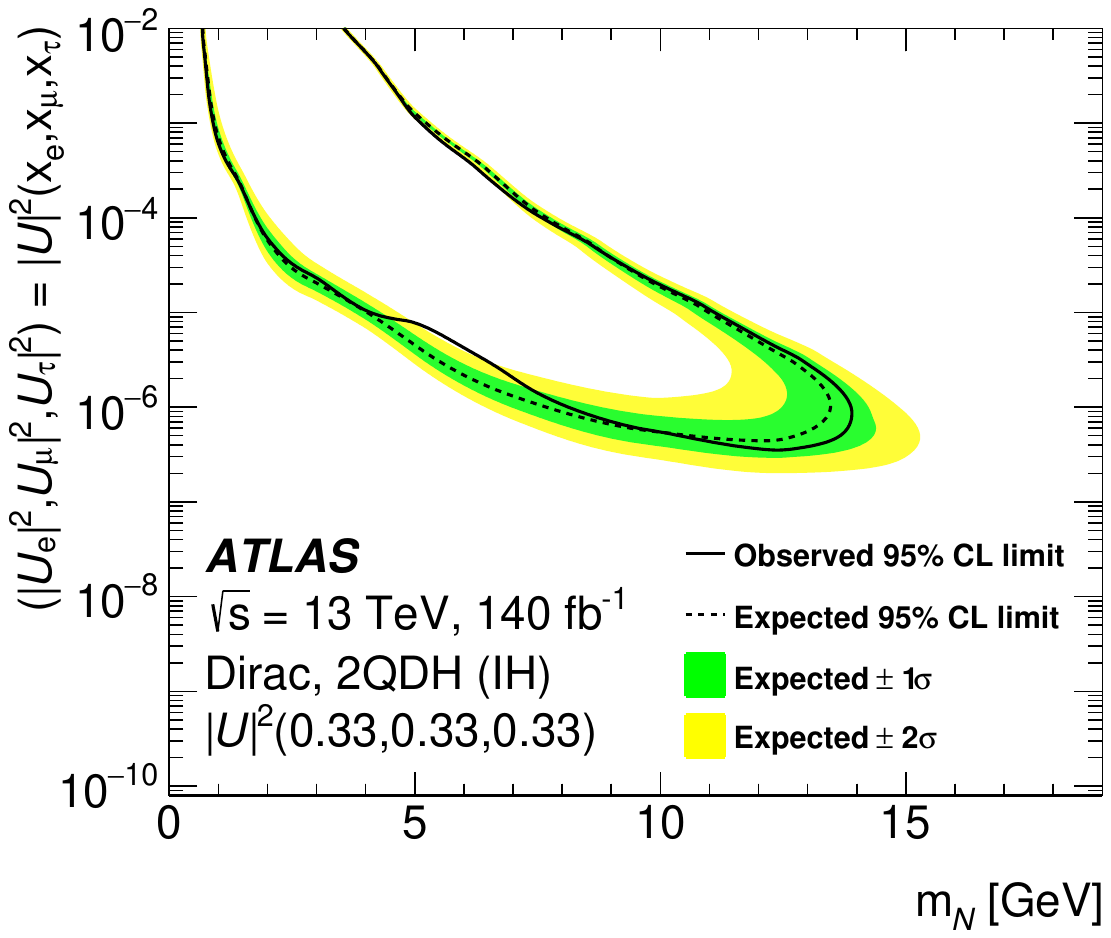}
\label{fig:limits:D_IH}
}
\caption{Expected and observed 95\% CL limits on $|U_{\alpha}|$ vs. $m_{\mathcal{N}}$ in the Dirac-limit case, with inner green and outer yellow bands showing the one and two standard deviation ($\sigma$) spreads for the expected limits. \protect\subref{fig:limits:D_SF_EL} 1SFH scenario with electron-only mixing, \protect\subref{fig:limits:D_SF_MU} 1SFH scenario with muon-only mixing, \protect\subref{fig:limits:D_NH} 2QDH scenario with normal (NH) mass hierarchy, \protect\subref{fig:limits:D_IH} 2QDH scenario with inverted mass hierarchy (IH). The parameters corresponding to the area within the contour are excluded.
}
\label{fig:limits:D}
\end{figure}

\begin{figure}[htbp]
\centering
\subfloat[]{
\includegraphics[width=0.45\textwidth]{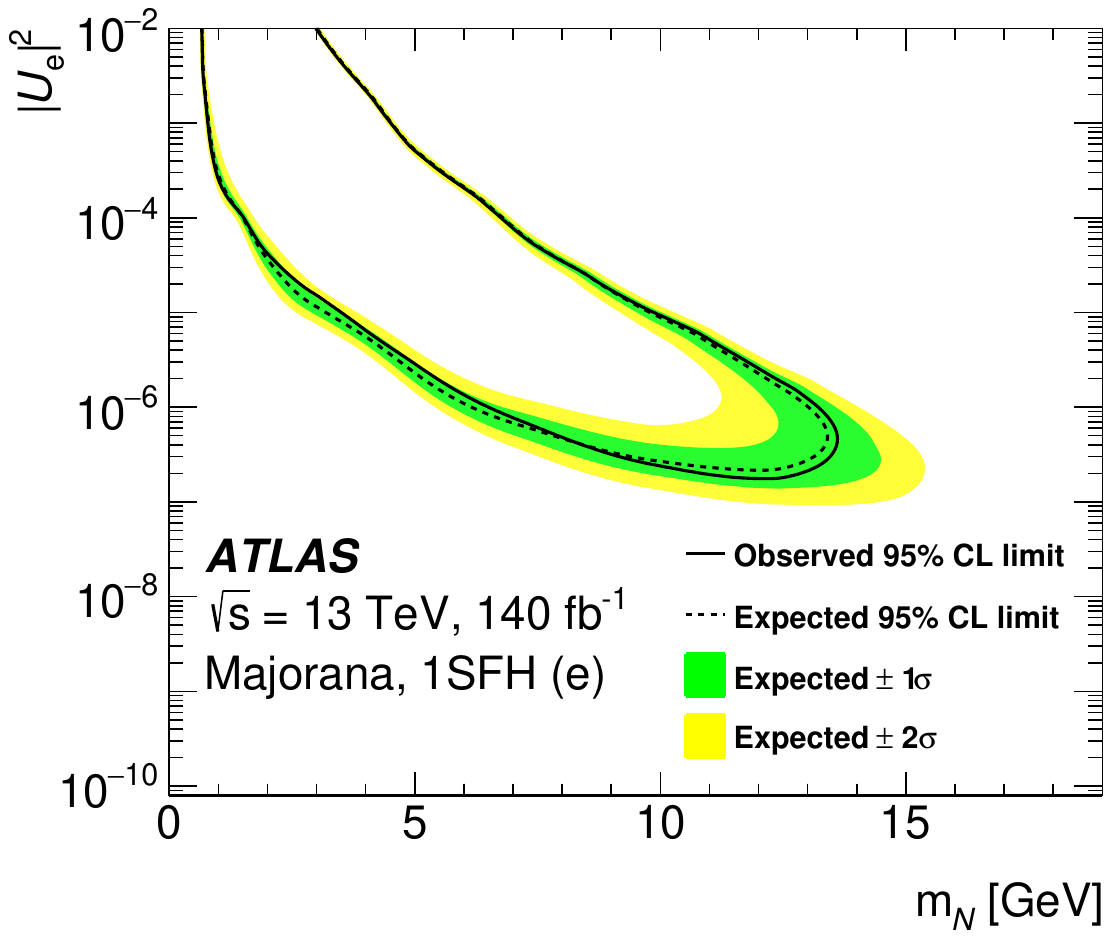}
\label{fig:limits:M_SF_EL}
}
\subfloat[]{
\includegraphics[width=0.45\textwidth]{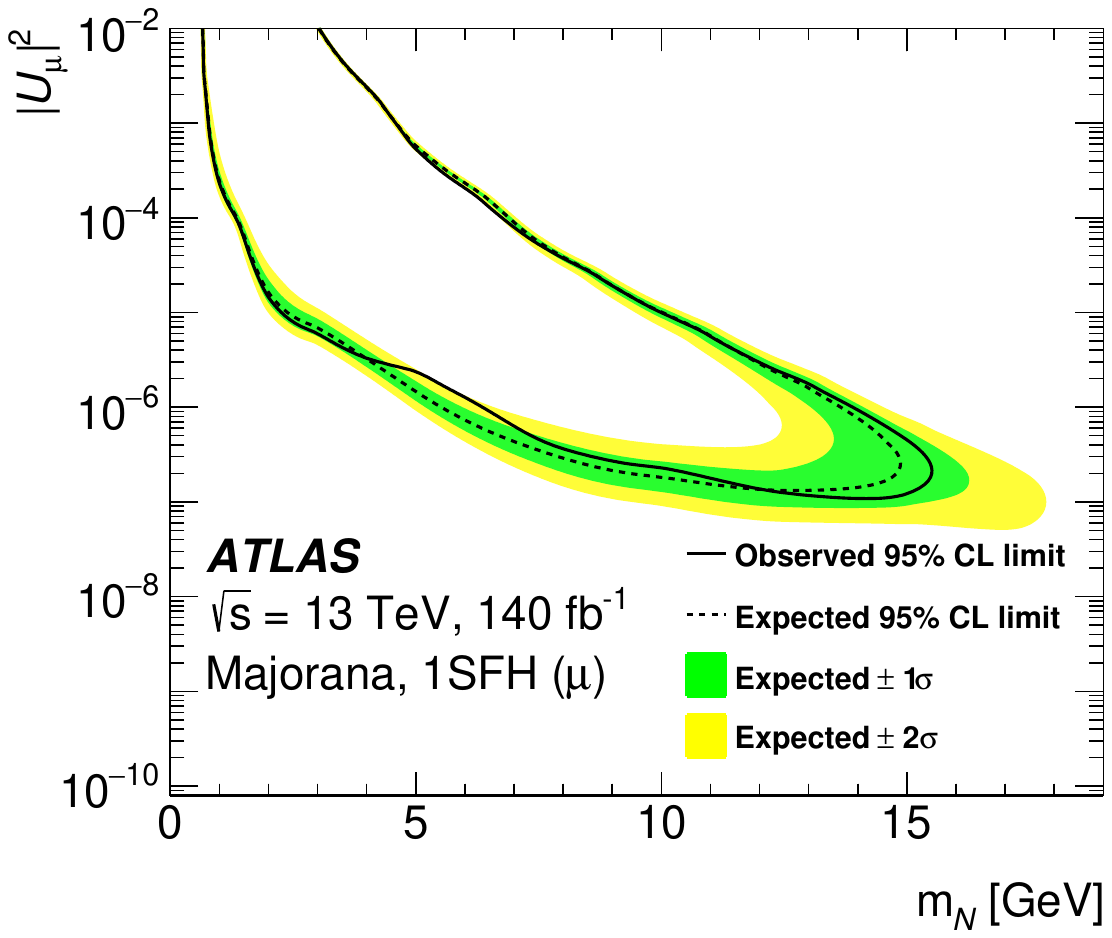}
\label{fig:limits:M_SF_MU}
}\\
\subfloat[]{
\includegraphics[width=0.45\textwidth]{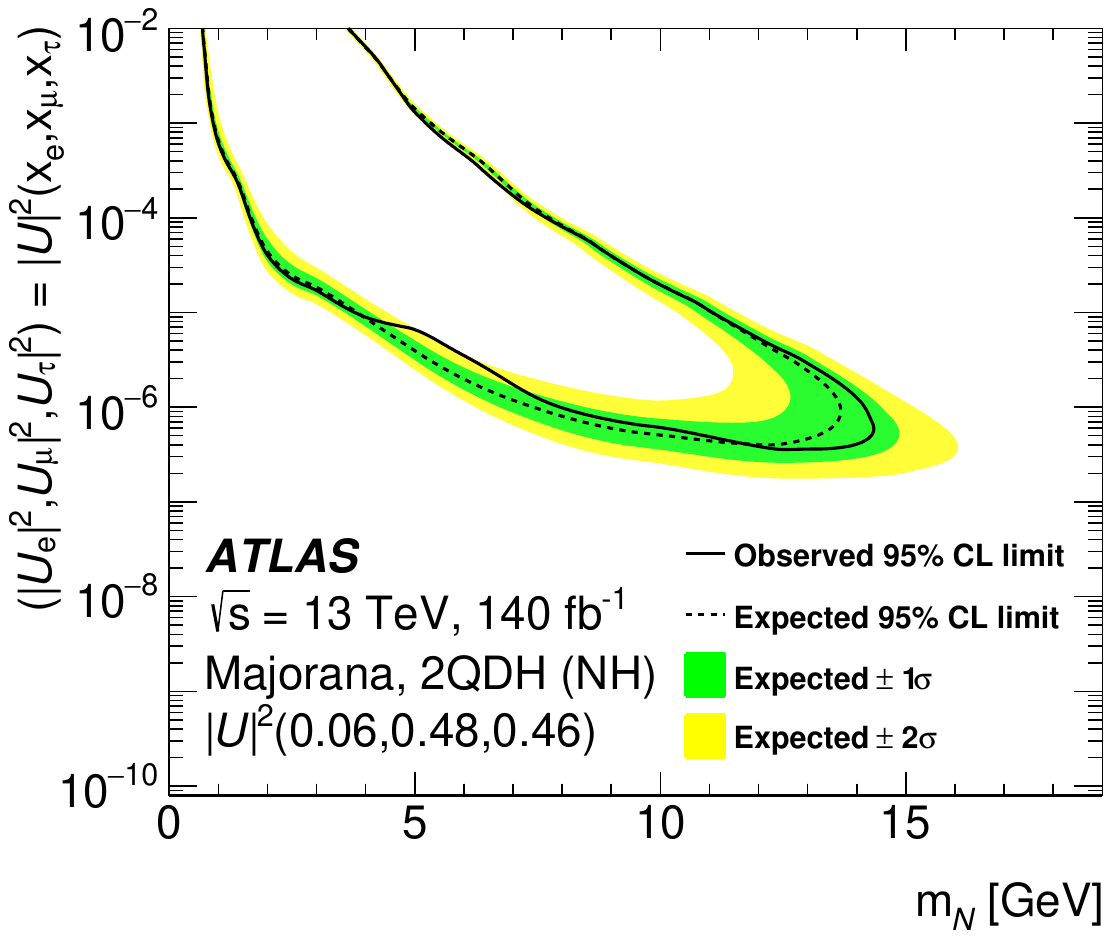}
\label{fig:limits:M_NH}
}
\subfloat[]{
\includegraphics[width=0.45\textwidth]{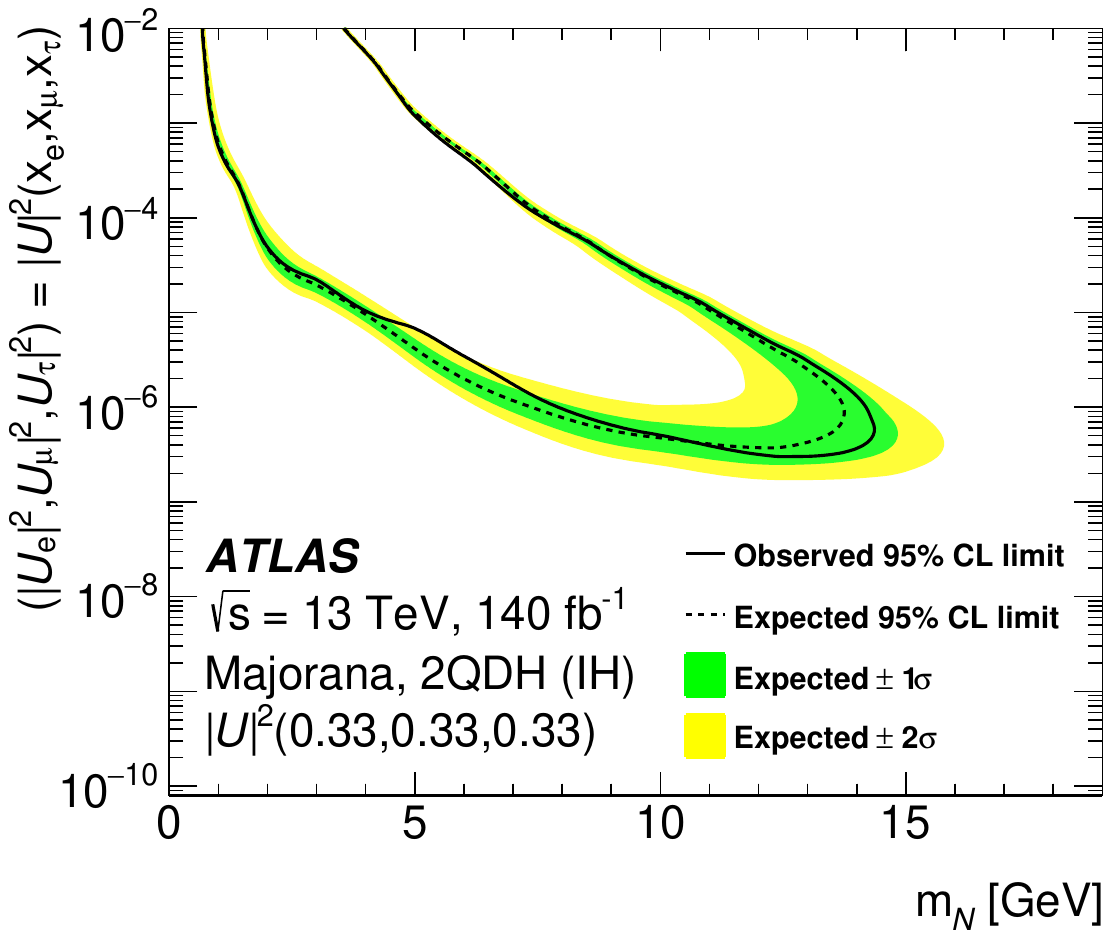}
\label{fig:limits:M_IH}
}
\caption{Expected and observed 95\% CL limits on $|U_{\alpha}|$ vs. $m_{\mathcal{N}}$ in the Majorana-limit case, with inner green and outer yellow bands showing the one and two standard deviation ($\sigma$) spreads for the expected limits. \protect\subref{fig:limits:M_SF_EL} 1SFH scenario with electron-only mixing, \protect\subref{fig:limits:M_SF_MU} 1SFH scenario with muon-only mixing, \protect\subref{fig:limits:M_NH} 2QDH scenario with normal (NH) mass hierarchy, \protect\subref{fig:limits:M_IH} 2QDH scenario with inverted mass hierarchy (IH). The parameters corresponding to the area within the contour are excluded.
}
\label{fig:limits:M}
\end{figure}

Limits are derived using the CL$_s$ prescription~\cite{Read:2002hq}, using the asymptotic formulae~\cite{Cowan:2010js}.
The upper limits on the signal strength of the different signal hypotheses are found to be in agreement within uncertainties with the results of hypothesis tests performed using $10^{4}$ pseudo-experiments for each signal parameter. The largest impact results in a shift in the limit contour of 0.25~\GeV in $m_\myhnl$ and $0.4\times10^{-7}$ in $|U|^2$.
Experimental, background-modelling and signal-modelling systematic uncertainties are included. The experimental uncertainties in the signal efficiency  are correlated with those on the background prediction.
The 95\% CL exclusion contours for the Dirac HNL are reported in Figure \ref{fig:limits:D}, and in Figure \ref{fig:limits:M} for the Majorana HNL\@. The feature in the observed 95\% CL limit contour at around 5~\GeV in $m_{\mathcal{N}}$ for the 1SFH model with muon-only coupling, as well as in both the 2QDH scenarios is related to the small signal-like excess of data events observed in the $\mu-\ell\ell$ SR for $m_{\text{HNL}} = 5~\GeV$, as shown in Figure \ref{fig:datamc:SRuxx}. The exclusion contours obtained for Majorana HNLs are wider compared to the ones obtained for the Dirac HNLs.
For the 1SFH model, a Majorana HNL has twice as many decay options, LNV as well as LNC, and thus has a shorter lifetime for a given mass and coupling value than a Dirac HNL.
Due to the relationship between decay position and efficiency as shown in Figure~\ref{fig:reco:DVeff}, this translates to higher sensitivity to Majorana HNLs.

The sensitivity of the search is limited by the available statistical precision in data.
At larger values of the coupling $|\Ua|^{2}$, the sensitivity is limited by the requirement $r_{\text{DV}} > \SI{4}{\mm}$.
For small values of $|U_\alpha|^{2}$, which correspond to large HNL lifetimes, the sensitivity is limited by the dimensions of the ATLAS inner detector and the requirement $r_{\text{DV}} < \SI{300}{\mm}$.
The sensitivity at larger $m_\mathcal{N}$ is limited mostly by the data statistics and by the precision on the estimate of the fake-leptons background. At low values of $m_\mathcal{N}$, the sensitivity is limited by the requirement $m_{\text{DV}} > 0.5\,\GeV$.
Stronger constraints are set on the parameters of models
with large values of $|U_{\mu}|^2$, since muons have a higher efficiency and lower background than electrons.

Compared with the previous search, which was performed with the same dataset but a previous version of the software to reconstruct large-radius tracks and secondary vertices~\cite{EXOT-2019-29}, this search extends the sensitivity to an HNL with muon-only coupling within the 1SFH scenario to larger values of $m_{\mathcal{N}}$ up to 14.5 \GeV, as well as to lower $|U_{\mu}|^2$ with respect to $|U_e|^2$ for $m_{\mathcal{N}}$ in the range of 4 -- 10~\GeV. The sensitivity is also improved for the 2QDH models, with only minor improvements for the 1SFH electron-only coupling case. Large gains in the sensitivity are obtained at low mass values, where contours are expanded to larger coupling values for low $m_\mathcal{N}$, bridging the gap with existing exclusion contours~\cite{Fernandez-Martinez:2023phj}.
The improvement arises from the change in the analysis strategy, the relaxation of selections that were previously necessary, along with the estimate of the HF-hadrons-production background. This estimate is now possible thanks to the introduction of the LRT reconstruction in the main ATLAS reconstruction workflow~\cite{IDTR-2022-04}, and the availability of these tracks in all the simulated samples.
As described, the semi-leptonic channels contribute to the signal sensitivity only for $m_{\mathcal{N}} \leq 3$~\GeV, where the limits are comparable to the leptonic case.

\FloatBarrier


%
\section{Conclusion}
\label{sec:conclusion}

A search for a long-lived heavy neutral leptons using 140 fb$^{-1}$ of proton--proton collision data collected with the ATLAS detector at the LHC is reported.
The search considers HNLs produced in $W$-boson decays and decaying to two leptons and a neutrino or to a lepton and two quarks, producing a pion.
Six leptonic and four semi-leptonic channels are considered, contributing to 1SFH($e$), 1SFH($\mu$), and 2QDH interpretations.
No significant excess is observed beyond the predicted background in any of the signal regions and limits are placed at a 95\% confidence level on the coupling $|\Ua|^2$ as a function of $m_{\mathcal{N}}$.
The search presented here places constraints on the production of HNLs in the mass range 0.5 $ < m_{\mathcal{N}} < $ 16 \GeV.
The sensitivity at low $m_{\mathcal{N}}$ values has improved compared with the previous ATLAS search~\cite{EXOT-2019-29} despite using the same dataset, due to improvements in the event reconstruction and the use of more sophisticated analysis techniques.


%
%
%
%
%
%
%
%

%
%
%
%
%
\section*{Acknowledgements}
%

%

%
%

%
%

We thank CERN for the very successful operation of the LHC and its injectors, as well as the support staff at
CERN and at our institutions worldwide without whom ATLAS could not be operated efficiently.

The crucial computing support from all WLCG partners is acknowledged gratefully, in particular from CERN, the ATLAS Tier-1 facilities at TRIUMF/SFU (Canada), NDGF (Denmark, Norway, Sweden), CC-IN2P3 (France), KIT/GridKA (Germany), INFN-CNAF (Italy), NL-T1 (Netherlands), PIC (Spain), RAL (UK) and BNL (USA), the Tier-2 facilities worldwide and large non-WLCG resource providers. Major contributors of computing resources are listed in Ref.~\cite{ATL-SOFT-PUB-2025-001}.

We gratefully acknowledge the support of ANPCyT, Argentina; YerPhI, Armenia; ARC, Australia; BMWFW and FWF, Austria; ANAS, Azerbaijan; CNPq and FAPESP, Brazil; NSERC, NRC and CFI, Canada; CERN; ANID, Chile; CAS, MOST and NSFC, China; Minciencias, Colombia; MEYS CR, Czech Republic; DNRF and DNSRC, Denmark; IN2P3-CNRS and CEA-DRF/IRFU, France; SRNSFG, Georgia; BMBF, HGF and MPG, Germany; GSRI, Greece; RGC and Hong Kong SAR, China; ICHEP and Academy of Sciences and Humanities, Israel; INFN, Italy; MEXT and JSPS, Japan; CNRST, Morocco; NWO, Netherlands; RCN, Norway; MNiSW, Poland; FCT, Portugal; MNE/IFA, Romania; MSTDI, Serbia; MSSR, Slovakia; ARIS and MVZI, Slovenia; DSI/NRF, South Africa; MICIU/AEI, Spain; SRC and Wallenberg Foundation, Sweden; SERI, SNSF and Cantons of Bern and Geneva, Switzerland; NSTC, Taipei; TENMAK, T\"urkiye; STFC/UKRI, United Kingdom; DOE and NSF, United States of America.

Individual groups and members have received support from BCKDF, CANARIE, CRC and DRAC, Canada; CERN-CZ, FORTE and PRIMUS, Czech Republic; COST, ERC, ERDF, Horizon 2020, ICSC-NextGenerationEU and Marie Sk{\l}odowska-Curie Actions, European Union; Investissements d'Avenir Labex, Investissements d'Avenir Idex and ANR, France; DFG and AvH Foundation, Germany; Herakleitos, Thales and Aristeia programmes co-financed by EU-ESF and the Greek NSRF, Greece; BSF-NSF and MINERVA, Israel; NCN and NAWA, Poland; La Caixa Banking Foundation, CERCA Programme Generalitat de Catalunya and PROMETEO and GenT Programmes Generalitat Valenciana, Spain; G\"{o}ran Gustafssons Stiftelse, Sweden; The Royal Society and Leverhulme Trust, United Kingdom.

In addition, individual members wish to acknowledge support from Armenia: Yerevan Physics Institute (FAPERJ); CERN: European Organization for Nuclear Research (CERN DOCT); Chile: Agencia Nacional de Investigaci\'on y Desarrollo (FONDECYT 1230812, FONDECYT 1240864); China: Chinese Ministry of Science and Technology (MOST-2023YFA1605700, MOST-2023YFA1609300), National Natural Science Foundation of China (NSFC - 12175119, NSFC 12275265); Czech Republic: Czech Science Foundation (GACR - 24-11373S), Ministry of Education Youth and Sports (ERC-CZ-LL2327, FORTE CZ.02.01.01/00/22\_008/0004632), PRIMUS Research Programme (PRIMUS/21/SCI/017); EU: H2020 European Research Council (ERC - 101002463); European Union: European Research Council (BARD No. 101116429, ERC - 948254, ERC 101089007), European Regional Development Fund (SMASH COFUND 101081355, SLO ERDF), Horizon 2020 Framework Programme (MUCCA - CHIST-ERA-19-XAI-00), European Union, Future Artificial Intelligence Research (FAIR-NextGenerationEU PE00000013), Horizon 2020 (EuroHPC - EHPC-DEV-2024D11-051), Italian Center for High Performance Computing, Big Data and Quantum Computing (ICSC, NextGenerationEU); France: Agence Nationale de la Recherche (ANR-21-CE31-0022, ANR-22-EDIR-0002); Germany: Baden-Württemberg Stiftung (BW Stiftung-Postdoc Eliteprogramme), Deutsche Forschungsgemeinschaft (DFG - 469666862, DFG - CR 312/5-2); China: Research Grants Council (GRF); Italy: Istituto Nazionale di Fisica Nucleare (ICSC, NextGenerationEU), Ministero dell'Università e della Ricerca (NextGenEU I53D23000820006 M4C2.1.1); Japan: Japan Society for the Promotion of Science (JSPS KAKENHI JP22H01227, JSPS KAKENHI JP22H04944, JSPS KAKENHI JP22KK0227, JSPS KAKENHI JP23KK0245, JSPS KAKENHI JP24K23939); Norway: Research Council of Norway (RCN-314472); Poland: Ministry of Science and Higher Education (IDUB AGH, POB8, D4 no 9722), Polish National Science Centre (NCN 2021/42/E/ST2/00350, NCN OPUS 2023/51/B/ST2/02507, NCN OPUS nr 2022/47/B/ST2/03059, NCN UMO-2019/34/E/ST2/00393, UMO-2022/47/O/ST2/00148, UMO-2023/49/B/ST2/04085, UMO-2023/51/B/ST2/00920, UMO-2024/53/N/ST2/00869); Portugal: Foundation for Science and Technology (FCT); Spain: Generalitat Valenciana (Artemisa, FEDER, IDIFEDER/2018/048), Ministry of Science and Innovation (MCIN \& NextGenEU PCI2022-135018-2, MICIN \& FEDER PID2021-125273NB, RYC2019-028510-I, RYC2020-030254-I, RYC2021-031273-I, RYC2022-038164-I); Sweden: Carl Trygger Foundation (Carl Trygger Foundation CTS 22:2312), Swedish Research Council (Swedish Research Council 2023-04654, VR 2021-03651, VR 2022-03845, VR 2022-04683, VR 2023-03403, VR 2024-05451), Knut and Alice Wallenberg Foundation (KAW 2018.0458, KAW 2022.0358, KAW 2023.0366); Switzerland: Swiss National Science Foundation (SNSF - PCEFP2\_194658); United Kingdom: Leverhulme Trust (Leverhulme Trust RPG-2020-004), Royal Society (NIF-R1-231091); United States of America: U.S. Department of Energy (ECA DE-AC02-76SF00515), Neubauer Family Foundation.

%
%


%
%
%
%
%
%

%
%

%
%
%
\printbibliography
\clearpage
 
\begin{flushleft}
\hypersetup{urlcolor=black}
{\Large The ATLAS Collaboration}

\bigskip

\AtlasOrcid[0000-0002-6665-4934]{G.~Aad}$^\textrm{\scriptsize 104}$,
\AtlasOrcid[0000-0001-7616-1554]{E.~Aakvaag}$^\textrm{\scriptsize 17}$,
\AtlasOrcid[0000-0002-5888-2734]{B.~Abbott}$^\textrm{\scriptsize 123}$,
\AtlasOrcid[0000-0002-0287-5869]{S.~Abdelhameed}$^\textrm{\scriptsize 119a}$,
\AtlasOrcid[0000-0002-1002-1652]{K.~Abeling}$^\textrm{\scriptsize 55}$,
\AtlasOrcid[0000-0001-5763-2760]{N.J.~Abicht}$^\textrm{\scriptsize 49}$,
\AtlasOrcid[0000-0002-8496-9294]{S.H.~Abidi}$^\textrm{\scriptsize 30}$,
\AtlasOrcid[0009-0003-6578-220X]{M.~Aboelela}$^\textrm{\scriptsize 45}$,
\AtlasOrcid[0000-0002-9987-2292]{A.~Aboulhorma}$^\textrm{\scriptsize 36e}$,
\AtlasOrcid[0000-0001-5329-6640]{H.~Abramowicz}$^\textrm{\scriptsize 157}$
% ...
\bigskip
\\

% ...
$^{17}$Department for Physics and Technology, University of Bergen, Bergen; Norway.\\
% ...
$^{30}$Physics Department, Brookhaven National Laboratory, Upton NY; United States of America.\\
% ...
$^{36}$$^{(a)}$Facult\'e des Sciences Ain Chock, Universit\'e Hassan II de Casablanca;$^{(b)}$Facult\'{e} des Sciences, Universit\'{e} Ibn-Tofail, K\'{e}nitra;$^{(c)}$Facult\'e des Sciences Semlalia, Universit\'e Cadi Ayyad, LPHEA-Marrakech;$^{(d)}$LPMR, Facult\'e des Sciences, Universit\'e Mohamed Premier, Oujda;$^{(e)}$Facult\'e des sciences, Universit\'e Mohammed V, Rabat;$^{(f)}$Institute of Applied Physics, Mohammed VI Polytechnic University, Ben Guerir; Morocco.\\
% ...
$^{45}$Physics Department, Southern Methodist University, Dallas TX; United States of America.\\
% ...
$^{49}$Fakult\"{a}t Physik , Technische Universit{\"a}t Dortmund, Dortmund; Germany.\\
% ...
$^{55}$II. Physikalisches Institut, Georg-August-Universit\"{a}t G\"ottingen, G\"ottingen; Germany.\\
% ...
$^{104}$CPPM, Aix-Marseille Universit\'e, CNRS/IN2P3, Marseille; France.\\
% ...
$^{119}$$^{(a)}$New York University Abu Dhabi, Abu Dhabi;$^{(b)}$United Arab Emirates University, Al Ain; United Arab Emirates.\\
% ...
$^{123}$Homer L. Dodge Department of Physics and Astronomy, University of Oklahoma, Norman OK; United States of America.\\
% ...
$^{157}$Raymond and Beverly Sackler School of Physics and Astronomy, Tel Aviv University, Tel Aviv; Israel.\\
% ...
$^{a}$ Also at Affiliated with an institute formerly covered by a cooperation agreement with CERN.\\
$^{b}$ Also at An-Najah National University, Nablus; Palestine.\\
$^{c}$ Also at Borough of Manhattan Community College, City University of New York, New York NY; United States of America.\\
$^{d}$ Also at Center for Interdisciplinary Research and Innovation (CIRI-AUTH), Thessaloniki; Greece.\\
$^{e}$ Also at Centre of Physics of the Universities of Minho and Porto (CF-UM-UP); Portugal.\\
$^{f}$ Also at CERN, Geneva; Switzerland.\\
% ...

\end{flushleft}


%

%
%

%

%
%
%

\end{document}